\documentclass[aps,prl,reprint,superscriptaddress,showpacs,longbibliography,footnoteinbib]{revtex4-2}
\usepackage[utf8]{inputenc}
\usepackage{xcolor}
\usepackage{float}
\usepackage{bm}
\usepackage{amsmath}
\usepackage{amssymb}
\usepackage{graphicx}
\usepackage{wasysym}
\usepackage{esint}
\PassOptionsToPackage{normalem}{ulem}
\usepackage{ulem}
\usepackage{soul}
\usepackage[unicode=true,
 bookmarks=false,
 breaklinks=false,pdfborder={0 0 1},backref=false,colorlinks=true]
 {hyperref}
\hypersetup{pdftitle={Title},
 plainpages=false,pdfpagelabels,linkcolor=blue,urlcolor=blue,citecolor=blue,pdfdisplaydoctitle=true,pdfduplex=DuplexFlipLongEdge}

\makeatletter

\newcommand{\lyxdot}{.}

\providecolor{lyxadded}{rgb}{0,0,1}
\providecolor{lyxdeleted}{rgb}{1,0,0}
\DeclareRobustCommand{\lyxsout}[1]{\ifx\\#1\else\sout{#1}\fi}

\usepackage{units}\usepackage{wasysym}

\definecolor{orange}{rgb}{0.50, 0.20, 0.0}

\newcommand{\beginsupplement}{%
	\setcounter{page}{1}
	 \renewcommand{\thepage}{SM - \arabic{page}}%
        \setcounter{table}{0}
        \renewcommand{\thetable}{S\arabic{table}}%
        \setcounter{figure}{0}
        \renewcommand{\thefigure}{S\arabic{figure}}%
        \setcounter{section}{0}
        \renewcommand{\thesection}{S\arabic{section}}%
        \setcounter{section}{0}
        \renewcommand{\thesection}{S\arabic{section}}%
        \setcounter{subsection}{0}
        \renewcommand{\thesubsection}{S\arabic{section}.\arabic{subsection}}%
        \setcounter{equation}{0}
        \renewcommand{\theequation}{S\arabic{equation}}%

     }

\usepackage{bm}
\usepackage{braket}

\makeatother

\begin{document}
\title{Doping-induced Quantum Anomalous Hall Crystals and Topological Domain
Walls}
\author{Miguel Gonçalves}
\affiliation{Theoretical Division T-4, Los Alamos National Laboratory, Los Alamos,
New Mexico 87545, USA}
\author{Shi-Zeng Lin}
\affiliation{Theoretical Division T-4, Los Alamos National Laboratory, Los Alamos,
New Mexico 87545, USA}
\affiliation{Center for Integrated Nanotechnologies (CINT), Los Alamos National
Laboratory, Los Alamos, New Mexico 87545, USA}
\begin{abstract}

Doping carriers into a correlated quantum ground state offers a promising route to generate new quantum states. The recent advent of moir\'{e} superlattices provided a versatile platform with great tunability to explore doping physics in systems with strong interplay between strong correlation and nontrivial topology. Here we study the effect of electron doping in the quantum anomalous Hall insulator realized in TMD  moir\'{e} superlatice at filling $\nu=1$, which can be described by the canonical Kane-Mele-Hubbard model. By solving the Kane-Mele-Hubbard model using an unrestricted
real-space Hartree-Fock method, we find that doping generates quantum anomalous
Hall crystals (QAHC) and topological domain walls. In the QAHC, the doping induces skyrmion spin textures, which hosts one or two electrons in each skyrmion as in-gap states. The skyrmions crystallize into a lattice, with the lattice parameter being tunable by the density of doped electrons. Remarkably, we find
that the QAHC can survive even in the limit of vanishing Kane-Mele
topological gap for a significant range of fillings. Furthermore, doping can also induce domain walls separating topologically distinct domains with different
electron densities, hosting chiral localized modes.
\end{abstract}
\maketitle
\noindent\begin{minipage}[t]{1\columnwidth}%
\global\long\def\ket#1{\left| #1\right\rangle }%

\global\long\def\bra#1{\left\langle #1 \right|}%

\global\long\def\kket#1{\left\Vert #1\right\rangle }%

\global\long\def\bbra#1{\left\langle #1\right\Vert }%

\global\long\def\braket#1#2{\left\langle #1\right. \left| #2 \right\rangle }%

\global\long\def\bbrakket#1#2{\left\langle #1\right. \left\Vert #2\right\rangle }%

\global\long\def\av#1{\left\langle #1 \right\rangle }%

\global\long\def\tr{\text{tr}}%

\global\long\def\Tr{\text{Tr}}%

\global\long\def\pd{\partial}%

\global\long\def\im{\text{Im}}%

\global\long\def\re{\text{Re}}%

\global\long\def\sgn{\text{sgn}}%

\global\long\def\Det{\text{Det}}%

\global\long\def\abs#1{\left|#1\right|}%

\global\long\def\up{\uparrow}%

\global\long\def\down{\downarrow}%

\global\long\def\vc#1{\mathbf{#1}}%

\global\long\def\bs#1{\boldsymbol{#1}}%

\global\long\def\t#1{\text{#1}}%
\end{minipage}

\vspace{-1cm}

\section{Introduction}

Moiré superlattices have emerged as a highly tunagble platform to explore strongly correlated topological quantum states. The kinetic energy of electrons can be tuned to be small in comparison to electron-electron interactions. 
As a consequence, a plethora of interaction-induced phases have been
experimentally observed in these systems, ranging from superconductivity, heavy fermion liquid, correlated insulator, Wigner crystal, the integer and fractional
quantum anomalous Hall effects \citep{Bistritzer_MacDonald_2011,Cao_Fatemi2018,Cao2018,Serlin_TschirhartYoung_2020,Sharpe_Goldhaber-Gordon_2019,Park2023,PhysRevX.13.031037,Xu_Liu_Shan_2020,Regan2020,Li_Li2021,Li_Jiang_Shen_Zhang2021,Zhao_Shen_Mak_Shan_2023,Cai2023,Zeng2023,park2024ferromagnetism,xu2024interplay,Park2021,xia2024unconventional,Guo_Pack_2024,Kang_Shen2024}.
A particular example of a very rich class of moiré systems is transition
metal dichalcogenides (TMDs). Some moiré TMDs can be described by triangular lattice Hubbard models with a trivial band topology
\citep{PhysRevLett.121.026402,Tang2020,PhysRevB.104.075150,PhysRevX.12.021031,doi:10.1126/sciadv.ade7701},
which has been verified by experimental observations \citep{Regan2020,Tang2020,Ciorciaro2023,doi:10.1126/science.adg4268,Tao2024}.
Of direct relevance to our work, homobilayer TMD moir\'{e} is also very appealing as it realizes generalized Kane-Mele-Hubbard
models \citep{PhysRevLett.122.086402,Devakul2021},
where the recent experimental observations of the integer
and fractional quantum anomalous Hall effects have been made \citep{Park2023,PhysRevX.13.031037,Cai2023,Zeng2023,Lu_Han_Yao_Fu_Ju_2024,park2024ferromagnetism,xu2024interplay}, inspired by model studies \cite{PhysRevLett.106.236802,PhysRevLett.106.236803,PhysRevLett.106.236804,sheng2011fractional,PhysRevX.1.021014,Xiao_Zhu_Ran_Nagaosa_Okamoto_2011} and particularly material-specific modeling \citep{PhysRevResearch.3.L032070,PhysRevB.108.085117}.

One unique advantage of moir\'{e} systems is that the carrier density can be controlled to fully fill or empty moir\'{e} bands by varying gate voltage. This immediately raises interesting questions about doping induced new physics, particularly doping around the correlation stabilized quantum many body states, and has attracted considerable attention recently. For instance, it was shown that doping electrons around a commensurate filling in twisted bilayer graphene stabilizes skyrmions \citep{doi:10.1126/sciadv.abf5299,Khalaf2022}, similar to the well known quantum Hall ferromagent for the Landau levels \citep{PhysRevB.47.16419}. It was further argued that the condensation of skyrmions can be the mechanism for the experimentally observed superconductivity in twisted bilayer graphene \citep{PhysRevLett.126.205701,doi:10.1126/sciadv.abf5299,PhysRevB.106.035421,PhysRevX.12.031020}. In TMD moir\'{e} superlattices, it was demonstrated that doping of carrier around half filling can stabilize spin polarons \citep{PhysRevB.107.224420,PhysRevLett.132.046501,Tao2024}, and can give rise to superconductivity \citep{PhysRevB.97.140507,Nazaryan_Fu_2024} or kinetic ferromagnetism \citep{Ciorciaro2023,PhysRevB.109.045144,PhysRevLett.132.046501}. Motivated by these exciting developments, we investigate the doping induced phases around a quantum anomalous Hall insulator (QAHI) in TMD homobilayer, such as twisted MoTe$_2$ and WSe$_2$, where the interplay between correlations and topology is essential. We unravel rich phases [see Fig.$\,$\ref{fig:1}(c) and Fig.$\,$\ref{fig:5}(a)], particularly the quantum anomalous Hall crystal (QAHC), stabilized by doping carriers into the correlated QAHI. Doping generates a skyrmion lattice with one or two electrons localized inside each skyrmion but with a different mechanism compared to that in quantum Hall ferromagnets. Furthermore, doping can create domain walls that separate topologically distinct regions with varying electron densities, hosting chiral localized modes.

\section{Main Results}

\begin{figure*}[t]
\centering{}\includegraphics[width=1\textwidth]{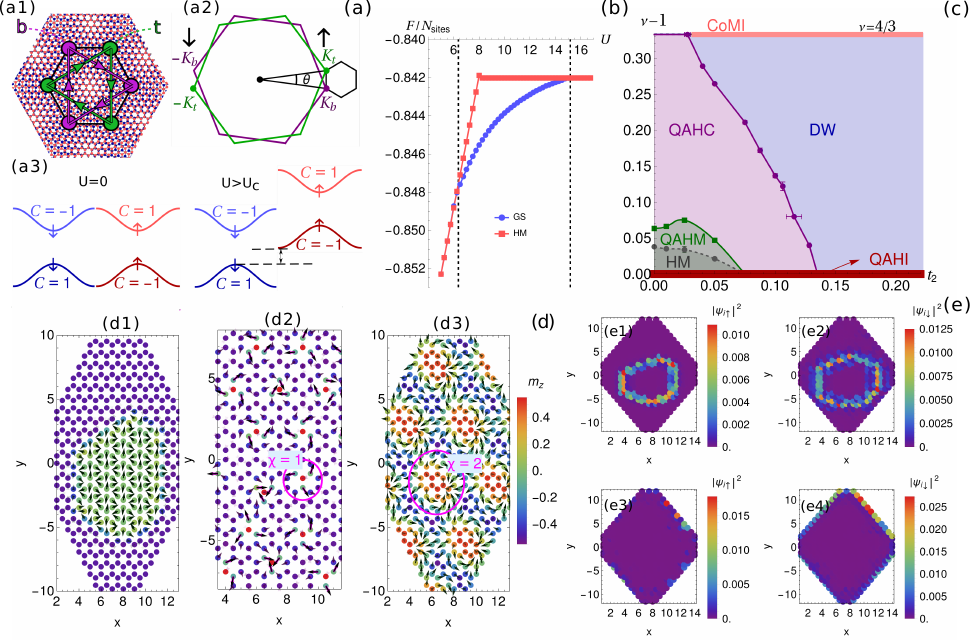}\caption{Main results. (a) Illustration of emergent Kane-Mele-Hubbard model
in TMD moiré superlattice. (b) Comparison of free energies of the half-metal
(HM) state and true ground state as a function of interaction strength
$U$, obtained for $t_{2}=0.2$, $L=24$ and $\delta_{e}=8$. (c) Phase
diagram for $U=8$ as a function of NNN hopping strength $t_{2}$
and filling fraction $\nu$. The different phases include a quantum
anomalous Hall insulator (QAHI), a quantum anomalous Hall crystal
(QAHC), a quantum anomalous Hall metal (QAHM), a topological domain
wall phase (DW), a half-metal (HM) and a coplanar magnetic insulator
(CoMI). {The critical points were estimated for different system sizes, as detailed in Supplementary Information Sec.$\,$\ref{sec:SUP_phase_boundary_details}, and the error bars correspond to the differences between the estimations for the two largest system sizes studied.} The lines connecting
the data points are guides to the eye. {Full and dashed lines indicate respectively first and second-order transitions.} 
(d) Site-resolved magnetization profile for $L=16$ and (d1) $t_{2}=0.125,\delta_{e}=26$
(DW); (d2) $t_{2}=0.1,\delta_{e}=26$ (QAHC); (d3) $t_{2}=0,\delta_{e}=18$
(QAHC). The color code denotes $m_{z}=\langle n_{i\protect\up}\rangle-\langle n_{i\protect\down}\rangle$,
while the arrows denote the vector ${\bf m}_{\perp}^{i}=(m_{x}^{i},m_{y}^{i})$
with magnitude encoded by the size and transparency of the arrow. (e) Site
and spin-resolved eigenstate at the Fermi energy for the parameters in (d1)
using periodic boundary conditions {[}(e1,e2){]} and the parameters
in (d2) using open boundary conditions {[}(e3,e4){]}. \label{fig:1}}
\end{figure*}

The low-energy electronic states in the twisted homobilayer TMD 
moiré such as MoTe$_2$ form a honeycomb lattice, with the two sublattices being layer polarized as illustrated in Fig.$\,$\ref{fig:1}(a1). Because of the strong spin orbit coupling, the valley and spin degrees of freedom are locked [see Fig.$\,$\ref{fig:1}(a2)], and in the following we use spin to denote both quantum numbers. The effective Hamiltonian is the Kane-Mele-Hubbard model \cite{PhysRevLett.122.086402}
\begin{equation}
\begin{aligned}H= & -t\sum_{\langle i,j\rangle,\sigma}c_{i,\sigma}^{\dagger}c_{j,\sigma}-t_{2}\sum_{\langle\langle i,j\rangle\rangle,\sigma}e^{-{\rm i}\sigma\phi_{ij}}c_{i,\sigma}^{\dagger}c_{j,\sigma}\\
 & +U\sum_{i}n_{i,\up}n_{i,\down}+V\sum_{\langle i,j\rangle,\sigma,\sigma'}n_{i\sigma}n_{j\sigma'}\,,
\end{aligned}
\label{eq:H}
\end{equation}
where $c_{i,\sigma}^{\dagger}$ creates an electron at site $i$ with
spin $\sigma$. The first term describes nearest-neighbor hoppings
on the honeycomb lattice. The second term describes next-nearest-neighbor (NNN) hoppings,
where $\phi_{ij}=\pm\pi/2$ and the sign is defined by the arrows
depicted in Fig.$\,$\ref{fig:1}(a1): $+$ ($-$) if the electron
hops along (against) the direction of the arrow. The final terms denote the onsite Hubbard and nearest-neighbor repulsive interaction. 
Throughout the manuscript, all results will be presented in units of the nearest-neighbor hopping
parameter $t$, which we set to $1$. We will also consider different electron fillings $\nu$, with $\nu=0$ and $\nu=4$ corresponding, respectively, to the fully empty and fully filled bands in Fig.$\,$\ref{fig:1}(a3). An electron filling $\nu$ here can be mapped to a doping of $\nu$ holes per moiré unit cell in TMD, by particle-hole transformation. We will focus on zero temperature.
The noninteracting band structure is illustrated in Fig.$\,$\ref{fig:1}(a3), where the opposite spins have the opposite Chern number and the system realizes a quantum spin Hall insulator at filling $\nu=2$ (equivalent to two holes per moiré unit cell).

The Hubbard
interactions can spontaneously split the degeneracy between spin up
and down sectors and give rise to the ferromagnetic QAHI at
filling $\nu=1$, corresponding to fully filling one of the bands, as shown in Fig.$\,$\ref{fig:1}(a3). 
The magnetization of the ground state is perpendicular to the moir\'{e} plane (see Supplementary Information,
Section \ref{sec:SUP_QAHI}) because of the Ising spin-orbit coupling. Below we study the novel phases induced by dopping around the $\nu=1$ ferromagnetic QAHI.

In Fig.$\,$\ref{fig:1}(b), we compare the free energies of the lowest energy homogeneous (translationally invariant) state and the true ground state. The former is a half-metal (HM), a spin polarized metal, which is the ground state below and above a critical interaction strength, consistent with recent exact diagonalization calculations in the large-$U$ regime
\citep{PhysRevB.107.L201109}. Interestingly, Fig.$\,$\ref{fig:1}(b) shows that there is a range of interaction strengths for which the ground state is not a HM. This
range is finite for all the fillings - from $\nu=1$ to $\nu=4/3$ - and intensities of
next-nearest-neighbor hopping strength/topological mass,
$t_{2}\in[0,0.2]$, studied in this paper.

The
main results are shown in Fig.$\,$\ref{fig:1}(c), where we selected
an interaction strength within the range where the ground state is
inhomogeneous and explored the possible phases in the plane of $t_{2}$
and filling $\nu$. For a large enough $t_{2}$, a topological domain
wall phase (DW) is stabilized for any finite electron doping away from $\nu=1$.
The domain wall separates two distinct magnetic domains: the ferromagnetic QAHI domain at filling $\nu=1$ with $|C|=1$ and
a topologically trivial coplanar magnetic insulator (CoMI) domain
at filling $\nu=4/3$, as depicted in Fig.$\,$\ref{fig:1}(d1). Although
the CoMI domain has $C=0$, its magnetization profile is non-trivial:
the unit cell becomes three times larger than the honeycomb unit cell
and the coplanar magnetization forms vortices. Due to an intricate
interplay between the non-trivial magnetization and the Kane-Mele
topological mass, these vortices are characterized by a winding number
determined by the sign of this mass, as we will detail below. Because
the CoMI and ferromagnetic domains are insulating and topologically
distinct, chiral localized modes naturally arise at the domain wall
as shown in Figs.$\,$\ref{fig:1}(e1,e2).

For a smaller $t_{2}$, there is a first-order phase transition into a QAHC with quantized Hall conductance $|\sigma_{xy}|=e^2/h$. In this phase, skyrmions are spontaneously
induced by doping the QAHI state. The skyrmions generate an emergent
magnetic field that couples with the orbital magnetization of the Chern band to minimize
energy. As a result, the Chern number of the filled Chern band determines
the sign of the skyrmion charge. An example
of the magnetization profile in this phase is shown in Fig.$\,$\ref{fig:1}(d2),
where a lattice of localized skyrmions spontaneously breaks the translational symmetry. Each skyrmion
accommodates exactly one electron, therefore the skyrmion crystal is also an electron crystal, in analogy to Wigner
crystals but with a quantized Hall conductance, as shown in the Supplementary
Information Sec.$\,$\ref{sec:SUP_different_phases}. The
lattice constant of the emergent skyrmion lattice is therefore determined by doping. It is important to note that for incommensurate fillings the crystal has imperfections: even though the skyrmions repel, favoring crystallization, it is not possible to form a perfect crystal. Nevertheless, these imperfections do not affect the quantization of Hall conductance. The QAHC has edge states exemplified in Figs.$\,$\ref{fig:1}(e3,e4),
with a uniform probability density for the background spin component
($\down$ in this example) and modulated by the skyrmion distribution
for the other spin component.

Surprisingly, the QAHC extends down to $t_{2}=0$, where the topological
mass vanishes. In this case, the
system is a ferromagnetic semimetal at $\nu=1$ due to interaction. However, at a finite doping, interactions stabilize
both spontaneous Chern gaps and skyrmions. Between the HM and the
QAHC there is a quantum anomalous Hall metal (QAHM) phase that has
a finite but non-quantized Hall conductance that perfectly correlates
with the total skyrmion charge, which is also smaller than the total number of doped electrons. This is also true for small but finite $t_{2}$,
showing that the creation of the QAHC does not require the existence
of a QAHI at $\nu=1$. The transition between the QAHM and QAHC phases is of first-order, as evidenced by the abrupt change in chemical potential exemplified in Supplementary Information Sec$\,$\ref{sec:SUP_phase_boundary_details}, Fig.$\,$\ref{fig:SUP_HM_to_QAHM_to_QAHC}(c). 

Finally, it is important to note that for $t_{2}=0$, each individual
skyrmion in the crystal has skyrmion charge $\chi=2$ and accommodates two electrons,
as shown in Fig.$\,$\ref{fig:1}(d3) (see also Supplementary Information
Sec.$\,$\ref{sec:SUP_different_phases}). Charge-2e skyrmions can
also be stabilized at finite $t_{2}$ by doping sufficiently away
from filling $\nu=1$ as we show in detail in the Supplementary
Information, Section \ref{sec:SUP_different_phases}.

\section{Origin of skyrmions}

\begin{figure}[t]
\centering{}\includegraphics[width=1\columnwidth]{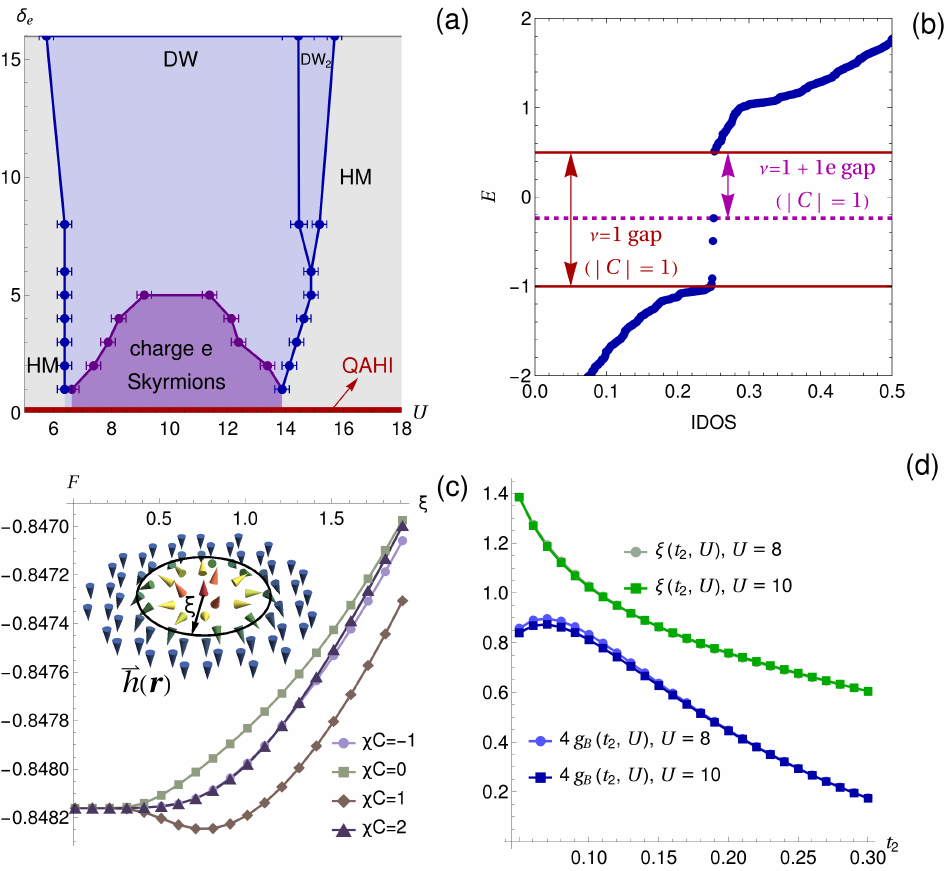}\caption{Origin of skyrmions. (a) Phase diagram as a function of interaction
$U$ and number of electrons $\delta_{e}$, for $t_{2}=0.2$. All transitions are first-order. The
error bars are the maximum between the difference in critical point
estimations for the two largest system sizes and the grid spacing
used for the calculations. (b) Eigenenergies as a function of integrated
density of states (IDOS) for $t_{2}=0.2$ and $U=7$. (c) Free energy of the system using the ansatz in Eq.$\,$\ref{eq:ansatz} for $L=20$,
$t_{2}=0.2$ and $U=8$. (d) The optimal skyrmion size $\xi$ that
minimizes the free energy and function $g_{B}(t_{2},U)$ defined in
Eq.$\,$\ref{eq:gB}, for different parameters. The curves are converged
to the thermodynamic limit.\label{fig:2}}
\end{figure}

We now uncover the mechanism behind the formation of skyrmions upon doing. For
a small, but finite, number of doped electrons, the ground state can
contain skyrmions in a wider range of $t_{2}$ than the one spanning
the QAHC in Fig.$\,$\ref{fig:1}(c). We show an example phase diagram
in the plane of $U$ and number of doped electrons $\delta_{e}$ in
Fig.$\,$\ref{fig:2}(a) for $t_{2}=0.2$, where only the DW phase
exists at finite density for $1<\nu<4/3$. In this phase diagram,
we observe that there is an interaction range where a skyrmion bound with an electron is generated by doping an electron. {This phase with electron-skyrmion bound states undergoes a first-order transition into the DW phase by varying $U$ or $\delta_e$. The number of skyrmions increases with $\delta_{e}$. The stability of this phase shrinks with increasing $\delta_{e}$, and 
there is a critical $\delta_{e}$ above which only the DW phase is
stable. The reason is that although the DW phase is energetically more favorable at finite electron density, its existence requires critical $\delta_e$}. We note that here we identified the ground state with the DW
phase whenever the different doped electrons cluster to form a domain.
In the Supplementary Information we detail the different structures
that can arise in the DW phase when doping a small number of electrons,
which do not necessarily have vanishing skyrmion charge. On a narrow
interaction range, right before we reach the HM phase at larger $U$,
a new metallic domain wall phase $\textrm{DW}_{2}$ arises.
{This phase is characterized by two domains with opposite magnetizations. However, unlike in the DW phase, one of the domains is metallic and no chiral edge modes arise at the domain wall. We analyze the $\textrm{DW}_{2}$ phase in more detail in Supplementary Information Sec.$\,$\ref{sec:SUP_different_phases}.} Here no QAHC is present since the formation of skyrmions is unstable at finite electron density compared to the DW phase.

In order to unravel the mechanism behind the formation of
skyrmions, in what follows we focus on the single-electron doping
problem, where a single skyrmion is formed. We can interpret the emergence
of the skyrmion as a magnetic impurity in the ferromagnetic background that is created in order to
save energy for the doped electron. Because
of this impurity, in-gap bound states are created as exemplified in
Fig.$\,$\ref{fig:2}(b). Filling these in-gap states saves energy compared
to overcoming the gap at filling $\nu=1$ to form a HM. However,
this does not explain why a magnetic impurity with unit skyrmion charge
is favorable compared to a simple spin-flip (polaron), or skyrmions
with larger charges.

To answer this question, we consider the ansatz described by $\langle n_{{\bf r},\sigma}\rangle=\frac{1}{4}(1+\sigma{\bf \hat{{\bf h}}}_{z}^{{\bf r}})$, $\langle c_{{\bf r},\up}^{\dagger}c_{{\bf r},\down}\rangle=\frac{1}{4}({\bf \hat{{\bf h}}}_{x}^{{\bf r}}+{\rm i}\hat{{\bf h}}_{y}^{{\bf r}})$, with ${\bf {\bf \hat{{\bf h}}}}^{{\bf r}}=(\sin\theta_{{\bf r}}\cos\phi_{{\bf r}},\sin\theta_{{\bf r}}\sin\phi_{{\bf r}},\cos\theta_{{\bf r}})$ and $\theta_{{\bf r}}=\pi[1-\arctan(|{\bf r}-{\bf r}_{0}|/\xi)]$, $\phi_{{\bf r}}=p\alpha$. In this ansatz, ${\bf r}_{0}$ is the position of the skyrmion's center, $\xi$
is the size of the skyrmion, $\alpha$ is the polar angle between the vector ${\bf r}-{\bf r}_{0}$ and the $x$-axis, and $p$ is an integer that determines the skyrmion charge $\chi$. This ansatz
gives rise to the exchange field texture illustrated in the inset
of Fig.$\,$\ref{fig:2}(c). Plugging it into Eq.$\,$\ref{eq:HMF}
(with $V=0$), we obtain

\begin{equation}
H_{MF}({\bf h})=H_{0}-\sum_{{\bf r}}{\bf h}({\bf r})\cdot{\bf s}_{{\bf r}}+\frac{U}{4}\sum_{{\bf r},\sigma}n_{{\bf r},\sigma}\,,
\label{eq:ansatz}
\end{equation}
where we have defined ${\bf h}({\bf r})=U{\bf {\bf \hat{{\bf h}}}}^{{\bf r}}/2$
and ${\bf s}^{{\bf r}}=\frac{1}{2}{\bf c}_{{\bf r}}^{\dagger}{\bf \sigma}{\bf c}_{{\bf r}}$,
with ${\bf c}_{{\bf r}}=[c_{{\bf r},\uparrow}\,,c_{{\bf r},\downarrow}]^{T}$
and ${\bf \sigma}$ being Pauli matrices. In the Supplementary
Information Sec.$\,$\ref{fig:SUP_validity_skyrmion_ansatz}, we show
that this ansatz captures very well the exact solution.
Using the ansatz, we show in Fig.$\,$\ref{fig:2}(c) an example of
the free energy as a function of $\xi$ in the regime where skyrmions
are stabilized, for different values of $\chi C$. Here  $C$ is the
ground state Chern number for $\delta_{e}=1$ and the skyrmion topological charge $\chi$
can be changed by varying $p$, with $\chi=(4\pi)^{-1}\int d^{2}{\bf r}\,{\bf m}_{{\bf r}}\cdot(\pd_{x}{\bf m}_{{\bf r}}\times\pd_{y}{\bf m}_{{\bf r}})\propto p$ [see Supplementary Information Sec.$\,$\ref{sec:SUP_Observables}]. Fig.$\,$\ref{fig:2}(c) shows that the free
energy is minimized at an optimal $\xi$ when $\chi C=1$, which means
that the Chern number determines the sign of the skyrmion charge.
We note that for the value of $\xi$ that minimizes the free energy,
the skyrmion is very localized and to a good approximation, the in-plane
components of the magnetization are non-vanishing only essentially
for the nearest-neighbors of site ${\bf r}_{0}$. Because of this,
and of $\mathcal{C}_{3}$ symmetry around ${\bf r}_{0}$, we have
$\chi = \mod(n,2)$. The skyrmion charge values shown in Fig.$\,$\ref{fig:2}(c)
therefore saturate all the possibilities, and no difference in the
free energy is observed between $\chi C=-1$ and $\chi C=2$. This
is no longer true at a larger $\xi$, where the skyrmion spreads over
further neighboring sites.

Based on these results, the mechanism for skyrmion formation arises
from a topological term in the free energy of the form $F_{B}=-C\chi g_{B}(t_{2},U)$, where $g_{B}$ is an unknown function of the model parameters. This
term is allowed by symmetry and arises from the coupling between the
orbital magnetization \cite{RevModPhys.82.1959} (proportional to $C$) of the Chern band and the emergent magnetic
field created by the skyrmion. \citep{dong2024chiral} In
the adiabatic limit, when the spin of the conduction electron follows the skyrmion texture, the emergent magnetic
field is $B_{S}\approx \phi_{0}\chi/(2\pi)\xi^2$,
with $\phi_{0}=hc/e$ the flux quantum and $\xi$ is the skyrmion size \cite{Batista_2016}. Away from the adiabatic limit, $B_S$ is expected to be reduced. Interestingly,
it is still possible to explicitly compute the function $g_{B}(t_{2},U)$
through 

\begin{equation}
g_{B}(t_{2},U)=|F_{\chi=1}(t_{2},U)-F_{\chi=-1}(t_{2},U)|.\label{eq:gB}
\end{equation}
where $F_{\chi}$ is the free energy obtained for skyrmion charge
$\chi$. For the Ising spin orbit coupling, $F_{\chi}$ depends on the sign of $\chi$ only through the topological term. As displayed in Fig.$\,$\ref{fig:2}(d), $|F_{B}|$ increases
with decreasing $t_{2}$, correlating with the increase in the skyrmion
size $\xi$. For a small enough
$t_{2}$ however, $|F_{B}|$ starts decreasing, which occurs approximately
when the HM becomes the ground state {[}see Fig.$\,$\ref{fig:1}(c){]}.
Finally, we note that since both the topological term and skyrmion
size $\xi$ decrease with $t_{2}$, the ansatz predicts
that the skyrmion becomes a polaron (a spin-flip magnetic impurity
with $\chi=0$) at a larger $t_{2}$. This is indeed what is observed
in the numerical calculation, as we detail in the Supplementary Information
Sec.$\,$\ref{sec:validity_ansatz}.

\section{Hall conductance in QAHC}

\begin{figure}[t]
\centering{}\includegraphics[width=1\columnwidth]{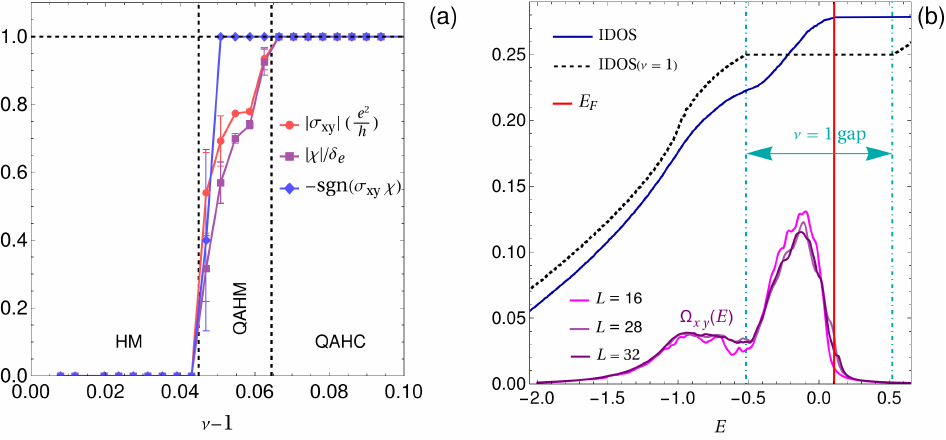}\caption{Hall conductance and Berry curvature. (a) Hall conductance $\sigma_{xy}$,
skyrmion charge $\chi$ and sign of $\sigma_{xy}\chi$ for $t_{2}=0$
and $L=16$. 
(b) Integrated density of states (IDOS) and energy-resolved Berry
curvature $\Omega(E)$ for $t_{2}=0.1$ and $\nu\approx1.112$. 
In red we show the Fermi energy
and in cyan we delimit the topological gap that exists at $\nu=1$.
We also plot the IDOS for $\nu=1$ (dashed black line) for comparison.
The results were averaged over the 10 ground states with
smaller free energies obtained from 250 mean-field calculations with
different fully random starting guesses. The error bars correspond to the standard deviation.
\label{fig:3}}
\end{figure}

In this section we discuss the origin of the quantized Hall conductance in the QAHC, as displayed in Fig.$\,$\ref{fig:3}(a). 
For a large enough $t_2$, when there is a sizable topological gap, one could argue that doping simply introduces in-gap skyrmion states that do not contribute to the Hall conductance, implying that skyrmions would not play a relevant role. However, the QAHC is very robust and extends down to vanishing $t_2$, where the $\nu=1$ topological gap vanishes. 
The reason for this robustness is that skyrmions always provide
a crucial contribution to the Hall response, even when the topological gap at $\nu=1$ is sizable. To show this, we examined the distribution of the Berry curvature as a function of energy, $\Omega(E)$ (Supplementary
Information Sec.$\,$\ref{sec:Berry-curvature-distribution} for details on the calculation). An example is shown in Fig.$\,$\ref{fig:3}(b). Upon doping the $\nu=1$ QAHI, when a sizable topological gap is present, skyrmions are created as in-gap states. However, the Berry curvature redistributes, acquiring a significant weight in these states as shown in Fig.$\,$\ref{fig:3}(b). This is compatible with the extended nature of these states, which is evidenced by the inverse participation ratio results shown in Supplementary Section Sec.$\,$\ref{sec:Berry-curvature-distribution}. While doping a finite number of electrons creates in-gap states exponentially localized around the skyrmions, at a finite density the skyrmion crystallization implies that these states become extended through hybridization with states in neighboring skyrmions.
For a small $t_{2}$ and higher dopings, the topological gap at $\nu=1$ is very small and it no longer makes sense to interpret skyrmions as arising from in-gap states upon doping. Instead, the Hall response becomes a direct consequence of the spontaneous crystallization of skyrmions, not requiring the existence of the QAHI at $\nu=1$. This is confirmed by the results of the Hall conductance at small or vanishing $t_2$. In this case, a critical doping with respect to $\nu=1$ is needed to induce both finite Hall response and total skyrmion charge, with these quantities perfectly correlating in the QAHM and QAHC phases, as shown in Fig.$\,$\ref{fig:3}(a) (see Supplementary
Information Sec.$\,$\ref{sec:SUP_Observables} for expression for
$\sigma_{xy}$).

It is important to note that even though the skyrmions repel
in the QAHC, which naturally favors crystallization, the skyrmion crystal is free to move as a whole in the clean limit.  
However, since impurities are always present in nature, it is expected that skyrmions are pinned by impurities in an experimental realization of the QAHC phase. As such, the QAHC is an insulator with quantized Hall conductance in the presence of a weak electric field. For a strong electric field,  the skyrmion crystal is driven into motion, rendering the system metallic and spoiling the quantization of the Hall conductance.

\section{Origin of domain wall state}

\begin{figure}[t]
\centering{}\includegraphics[width=1\columnwidth]{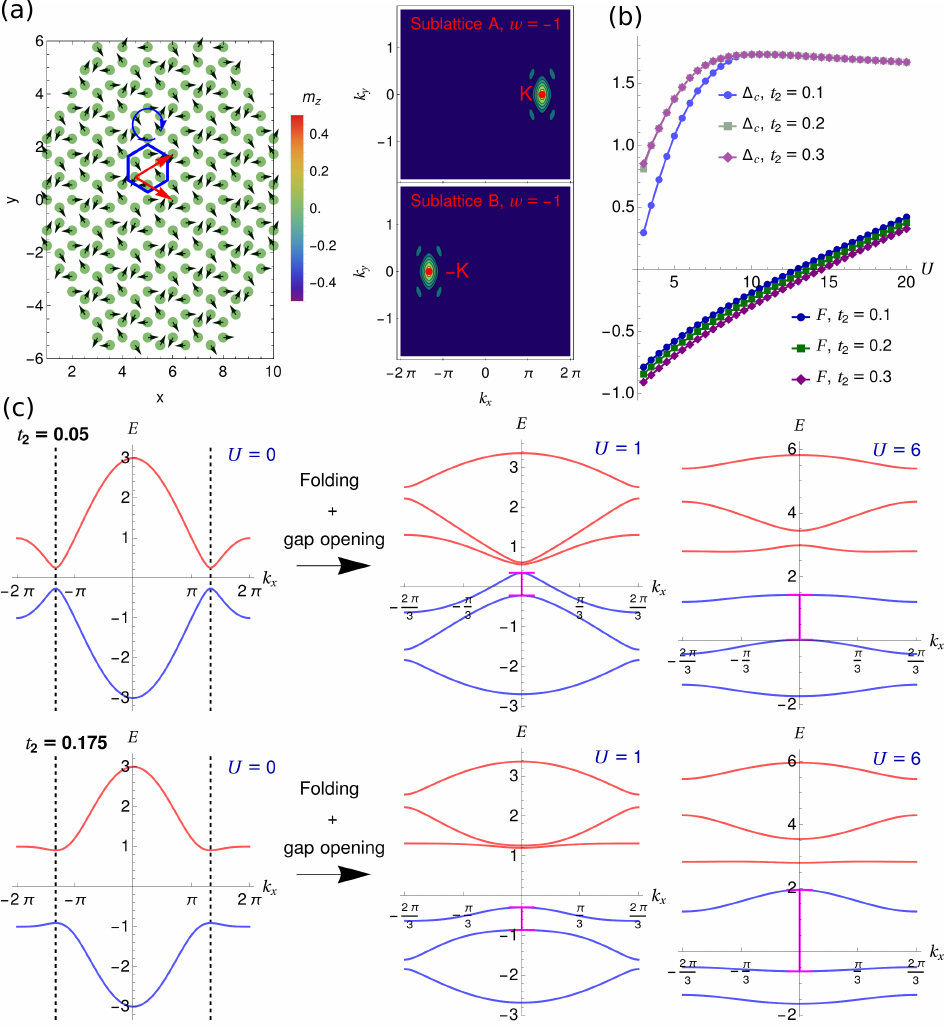}\caption{Origin of coplanar domain. (a) Magnetization profile at filling $\nu=4/3$
and $L=12$. The new unit cell is indicated in blue and the lattice
vectors in red. The blue arrow indicates the non-trivial vorticity
of the coplanar magnetization around the center of the unit cell.
The rightmost figures correspond to $F_{\up \down}({\bf k})=\sum_{{\bf r}_{m}}e^{-i{\bf k}\cdot{\bf r}_{m}}\langle c_{{\bf r}_{m},\protect\up}^{\dagger}c_{{\bf r}_{m},\protect\down}\rangle$,
where $m=\pm1$ for sublattices $A$ and $B$ respectively. (b) Free
energy $F$ and energy gap $\Delta_{c}$ of the CoMI ground state
for different model parameters. (c) Band structure for $M=0.285U$
(value chosen according to numerical results) and $k_{y}=0$, taking
$t_{2}=0.05$ (top row) and $t_{2}=0.175$ (bottom row), and with
$k_{x}$ running in the original (left) and folded (right)
Brillouin zones. \label{fig:4}}
\end{figure}

We now turn to understand the origin of the DW phase and its stability
at a larger $t_{2}$. The coplanar domain {[}see Fig.$\,$\ref{fig:1}(d1){]}
has a charge density $\langle n_{i}\rangle=2/3$ corresponding to
an electron filling $\nu=4/3$, in contrast with the ferromagnetic
domain that has $\langle n_{i}\rangle=1/2$ corresponding to filling
$\nu=1$ [Supplementary Information Sec.$\,$\ref{sec:SUP_different_phases}]. Both the $\nu=1$ ferromagnetic QAHI and the $\nu=4/3$ coplanar state are the most energeetically favorable states around $\nu=1$ fillings, and it is preferable for the system to have phase separation with coexisting $\nu=1$ and $\nu=4/3$ domains.  At $\nu=4/3$, the
coplanar domain fills the whole system as shown in Fig.$\,$\ref{fig:4}(a),
forming a topologically trivial gapped magnetic ground state. The
unit cell for this state triples in size compared to the original
unit cell, as shown in Fig.$\,$\ref{fig:4}(a). Furthermore, the
coplanar magnetization assumes the form ${\bf m}_{{\bf r}}=M[\cos\theta_{{\bf r}},\sin\theta_{{\bf r}},0]$
with a finite winding number $w=(2\pi)^{-1}\oint d^{2}{\bf r}\,{\bf \nabla}\theta_{{\bf r}}\cdot d{\bf r}$,
where the integral is performed in a closed contour connecting the
unit cell sites. Below, we will show that $w$ is fully determined
by the sign of the topological mass. 

In the Supplementary Information Sec.$\,$\ref{sec:SUP_coplanar_magnet},
we obtain the full band structure of the CoMI at $\nu=4/3$. In Fig.$\,$\ref{fig:4}(b),
we plot the free energy and energy gap for different parameters. The
free energy increases with $U$ and decreases with $t_{2}$, which
is consistent with the instability of the DW state at sufficiently
large $U$ {[}Fig.$\,$\ref{fig:1}(b){]} and small $t_{2}$ {[}Fig.$\,$\ref{fig:1}(c){]}.
This can be better understood by analysing the band structure. In
Fig.$\,$\ref{fig:4}(c) we show the band structure for different
$U$ and for two significantly different values of $t_{2}$. In the
lefmost figure, we plot the band structure for $U=0$ in the original
hexagonal Brillouin zone. Since the unit cell is tripled for the
CoMI, band folding occurs for $U=0$ and gaps are opened at the degeneracy
points for $U\neq0$. Note that for $U\neq0$, each shown band is
doubly degenerate as we will detail below. For a large value of
$t_{2}$, we can get very narrow bands around the Fermi level due
to the small dispersion around the Dirac points. Because of this,
it is possible to significantly decrease free energy by opening a
gap at $\nu=4/3$ {[}shown in magenta in Fig.$\,$\ref{fig:4}(c){]}.
For a smaller $t_{2}$, the folded energy bands are more dispersive,
and it is not possible to save as much energy by opening the gap.
Another interesting feature of this band structure is that a significantly
larger gap is opened around filling $\nu=4/3$ than around $\nu=8/3$.
In fact, for a small $U$, the gap around $\nu=8/3$ is essentially
suppressed. To understand the underlying mechanism, we derive
an approximate continuum model for the 8 central bands. 

In Fig.$\,$\ref{fig:4}(a-right), we show the Fourier transforms
of $\langle c_{{\bf r}_{m},\up}^{\dagger}c_{{\bf r}_{m},\down}\rangle$,
that peak at ${\bf K}$ or ${\bf K}'$ depending on the sublattice
and on $w$ (not shown). This gives rise to the following contribution
for the mean-field Hamiltonian, $Me^{-i\sigma wm\phi}\sum_{{\bf b},{\bf k}}c_{{\bf k},m,-\sigma}^{\dagger}c_{{\bf k}-\sigma wm({\bf K}+{\bf b}),m,\sigma}$,
where $M\propto U$, $m=\pm1$ respectively for sublattices $A$ and
$B$, ${\bf b}$ are the reciprocal lattice vectors and
$\phi=\pi/6$ is the angle difference between the magnetization vector
at different sublattices. Because low-energy physics is dominated by the states
around the Dirac points, we take ${\bf k}=-{\bf K},\ {\bf K}$ and consider
the first-order processes resulting from momentum transfers in this term.
From this we can derive the following low-energy continuum Hamiltonian
(see Supplementary Information Sec.$\,$\ref{sec:SUP_coplanar_magnet}
for details),

\begin{equation}
\begin{array}{c}
H=v_{f}\bar{\psi}(s_{x}\tau_{z}\pd_{x}+s_{y}\pd_{y})\psi+\lambda_{\textrm{SO}}\bar{\psi}s_{z}\tau_{z}\sigma_{z}\psi\\
+\frac{M}{2}\bar{\psi} \tau_{x}\Big(c_{\phi}\sigma_{x}-s_{\phi}\sigma_{y}\Big) \psi+\frac{wM}{2}\bar{\psi} s_{z}\tau_{y}\Big(c_{\phi}\sigma_{y}+s_{\phi}\sigma_{x}\Big)\psi
\end{array}\,,\label{eq:H_continuum}
\end{equation}
where ${\bf s},{\bf \tau},{\bf \sigma}$ are Pauli matrices acting
respectively on the sublattice, valley and spin subpaces,  $c_{\phi}=\cos(\phi)$,
$s_{\phi}=\sin(\phi)$ and $\lambda_{\textrm{SO}}=-3\sqrt{3}t_{2}$.
By rearranging ${\bf \psi}$ in the sub-blocks $(\psi_{A,{\bf K},\up},\psi_{B,{\bf K},\up},\psi_{A,{\bf K}',\down},\psi_{B,{\bf K}',\down})$
and $(\psi_{B,{\bf K},\down},\psi_{A,{\bf K},\down},\psi_{B,{\bf K}',\up},\psi_{A,{\bf K}',\up})$,
the Hamiltonian becomes block-diagonal with identical blocks, which
explains the double degeneracy of the band. Note that in contrast to the $U=0$
case, this is not a spin degeneracy. Defining $s_{w}=\sgn(w\lambda_{so})$,
we can easily derive that 

\begin{equation}
\begin{array}{c}
E({\bf q}=0)=\begin{cases}
-|\lambda_{\textrm{SO}}|-M\delta_{s_{w},1}\\
-|\lambda_{\textrm{SO}}|+M\delta_{s_{w},1}\\
|\lambda_{\textrm{SO}}|-M\delta_{s_{w},-1}\\
|\lambda_{\textrm{SO}}|+M\delta_{s_{w},-1}
\end{cases}\end{array}
\end{equation}
where ${\bf q}$ is measured from the Dirac points. This
implies that depending on the sign of $w\lambda_{\textrm{SO}}$, the gap can
open either at the lower- or higher-energy bands. In order to minimize
energy at filling $\nu=4/3$, the former case is more favorable. Therefore,
quite remarkably, the sign of the topological mass $\lambda_{\textrm{SO}}$
fixes the winding number $w$ of the coplanar domain. This phenomenon
is only possible when $\lambda_{\textrm{SO}}\neq0$. Note also that there is
no dependence on the sub-lattice angle difference $\phi$ (different
values simply define different $U(1)$ spin rotations, see Eq.$\,$\ref{eq:H_continuum}).
This is only an artifact of our first-order expansion to obtain the
continuum model. The full exact calculation has a $\phi$ dependence
(that manifests more significantly at larger $U$), setting $\phi=\pi/6$
as the value that minimizes free energy.

\section{Robustness of QAH and DW phases}

Even though the onsite Hubbard interaction is the dominant one in twisted
TMDs, longer-range interactions can also be quite significant. In what follows, we study the robustness of the
phases unveiled here against nearest-neighbor interactions. 

\begin{figure}[t]
\centering{}\includegraphics[width=1\columnwidth]{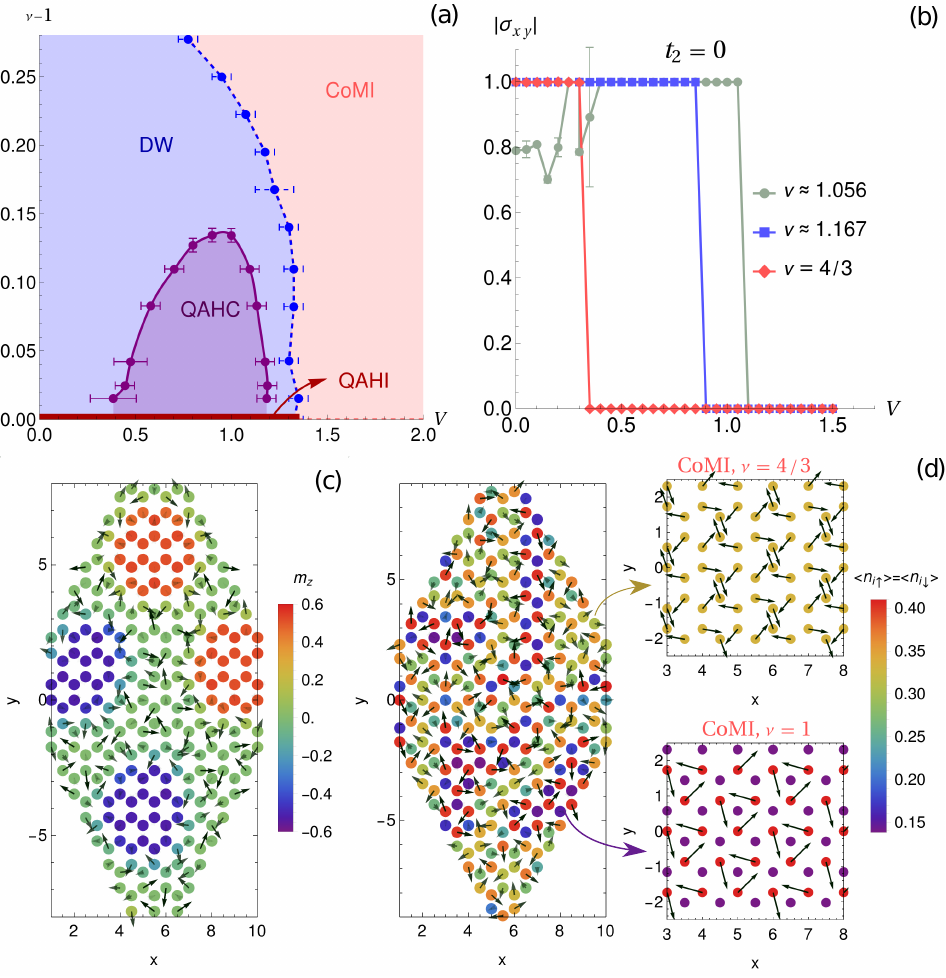}\caption{Robustness of QAHC and DW phases for a nonzero nearest-neighbor interaction $V$.
(a) Phase diagram for $U=8$ and $t_{2}=0.2$. Full and dashed lines denote respectively the  first and second-order transitions. The error bars are
the maximum between the difference in critical point estimations for
the two largest system sizes and the grid spacing used for the calculations.
(b) Hall conductance for $L=18$, $t_{2}=0$ and $U=8$. The error
bars are differences between the results for the 2 lowest free energy
ground states computed out of $1000$ different mean-field calculations.
(c) Magnetization profile in the DW phase for $L=12$, $V=1.05$
and $\nu\approx1+0.168$. (d) Magnetization profile in the CoMI ground state for
$L=12$, with (d1) $V=1.35$, $\nu\approx1+0.224$; (d2) $V=1,\nu=4/3$
and (d3) $V=1.75,\nu=1$. \label{fig:5}}
 
\end{figure}

We start by studying the stability of the DW phase. In Fig.$\,$\ref{fig:5}(a),
we vary $V$ for $t_{2}=0.2$ and $U=8$, where only the DW phase
exists at $V=0$ {[}see Fig.$\,$\ref{fig:1}(c){]}. We show that
the DW phase is robust up to a significantly large $V$, above which
the ground state becomes a CoMI, with $\text{\ensuremath{\langle m_{z}^{i}\rangle=0}}$.
Furthermore, there is a $V$-induced reentrant transition into the
QAHC, that is stabilized for intermediate $V$ and for $\nu$ close to
$\nu=1$. Two important comments are in order regarding the features
of the finite-$V$ DW phase. A finite $V$ favors a staggered density
between NN sites. Since the density is constant in the ferromagnetic
domain, it is energetically favorable to split it in smaller domains
to increase the number of NN with staggered density at the domain
walls. The size of the domains is therefore dictated by the competition
between $U$ and $V$. In particular, while the number of sites in
the domain wall for a single domain scale as $N_{S}^{DW}\propto L$,
it scales as $N_{S}^{DW}\propto N_{DW}L_{DW}\propto L^{2}$ for a
crystal of domains, where $N_{DW}\propto L^{2}$ is the number of
magnetic domains and $L_{DW}$ is the domain wall size of each domain,
that we assume to be $L$-independent. For a large enough $V$ it can
therefore be more favorable to create different domains of fixed size
than a single domain to maximize the energy gain at the domain walls.
An example ground state corroborating this picture is shown in Fig.$\,$\ref{fig:5}(c). 
Regarding the coplanar domain, for finite $V$ both the density and
magnitude of coplanar magnetization acquire a sub-lattice modulation,
which can also be seen in the example in Fig.$\,$\ref{fig:5}(c)
{[}see Supplementary information Sec.$\,$\ref{sec:SUP_different_phases}
for plots of the electron density{]}. 

As $V$ is increased further, the ferromagnetic domains vanishes smoothly through a second-order phase transition into the CoMI phase. The spin texture in the CoMI phase is also quite complex. At $\nu=1$ within this phase, a topologically trivial gapped coplanar domain
with strong sub-lattice density and coplanar magnetization modulation
is formed {[}see Fig.$\,$\ref{fig:5}(d){]}, while at $\nu=4/3$
the $V=0$ coplanar ground state is very stable in the studied range
of $V$. For $1<\nu<4/3$, domains that inherit from the $\nu=1$
and $\nu=4/3$ ground states are formed, as exemplified in Fig.$\,$\ref{fig:5}(d).
Since both these domains are topologically trivial, no edge states
arise at the domain walls, and the system is still a trivial gapped
insulator at these intermediate fillings.

Finally, we check the stability of the $t_{2}=0$ QAHC for a nonzero
$V$. Fig.$\,$\ref{fig:5}(b) shows that this phase can be stable
up to $V\approx1$ for the smallest filling fractions for which it
develops. The smallest skyrmions that form the crystal again accommodate
two electrons and have topological charge $\chi=2$ (not shown), as for $V=0$.
However, for larger fillings, the QAHC becomes less robust to $V$,
as shown in Fig.$\,$\ref{fig:5}(b).

\section{Discussion }

In this paper, we studied the ground state phase diagram arising from
doping a quantum anomalous Hall insulator. We found that depending
on doping and on the strength of interactions and non-interacting
topological mass, a quantum anomalous Hall crystal phase competes
with a topological domain wall phase, both spontaneously breaking
time-reversal and translational symmetry. Remarkably, topology plays
a crucial role in determining the nature of both phases.

Arguably the most important results are the robustness and tunability of the QAHC
phase for fillings significantly away from $\nu=1$ and down to
$t_{2}=0$, showing that a finite non-interacting topological mass
is not necessary for the spontaneous realization of the QAHC state.
In fact, the crystallization of skyrmions becomes the key ingredient
behind the QAHC. To our knowledge, this is the first
example of a non fine-tuned spontaneous QAHC arising from a topologically trivial
parent state. Previous examples of spontaneous QAHC effect arising
from trivial bands, also induced by noncoplanar spin textures, require specific fillings
where perfect Fermi surface nesting conditions are met \citep{PhysRevLett.101.156402,Li2012,PhysRevB.85.035414,PhysRevX.4.031040,PhysRevB.85.073103},
such as filling up to the Van Hove singularities. In contrast, in the present
case, the spontaneous skyrmion crystals exist for a significant range
of doping and are highly tunable, despite the presence of the tight-binding lattice. By varying both the model parameters and the filling fraction, it is possible to stabilize phases with charge-e and charge-2e skyrmions, or even with more complicated patterns such as stripes of skyrmions as we show in 
Supplementary Information Sec.$\,$\ref{sec:SUP_different_phases}.
These features also distinguish the QAHC phase
unveiled here from the anomalous Hall crystals recently proposed
in the literature \citep{zhou2023fractionalquantumanomaloushall,dong2023anomaloushallcrystalsrhombohedral,kwan2023moirefractionalcherninsulators,tan2024parentberrycurvatureideal,sheng2024quantumanomaloushallcrystal,Dong_Patri_Senthil_2024,PhysRevLett.132.236601}.

It is important to note that the DW phase discussed here is also different from previous examples.
It has previously been shown that doping a gapped correlated insulator can spontaneously induce
different domains, with the extra charge accommodated in topologically protected bound states in the domain walls \citep{LopezSancho2018,PhysRevB.108.045401,PhysRevLett.125.240601},
a mechanism well captured by the Jackiw-Rabbi model \citep{PhysRevD.13.3398}.
This mechanism can be activated due to the existence of an inhomogeneous
potential (e.g., a spatially varying substrate potential), which has
been proposed in twisted bilayer graphene \citep{PhysRevB.103.075122,PhysRevB.103.075423,PhysRevB.104.115404},
consistent with experimental observations \citep{Grover2022}. It
can also be at play at finite temperatures due to an increase in entropy from the topological localized modes that are located on the domain
wall \citep{PhysRevLett.128.156801}.
In contrast, in our example the DW phase is not induced by entropy. Furthermore, the CoMI domain accommodates most of the doped electrons, since it is very energetically favorable,
with the domain walls playing a sub-leading role.

We finally comment on the possible relevance of our results to moir\'{e} TMD, such as MoTe$_2$ and WSe$_2$. Starting from the QAHI at $\nu=1$, our results show that the gap survives over an extended region of doping. In the QAHC, the Hall conductance remains quantized against doping of electrons. In the DW phase, there coexist topological and trivial gapped phases corresponding, respectively, to $\nu=1$ and $\nu=4/3$. As we gradually dope the system away from $\nu=1$, as far as the topological domain percolates the whole system, the Hall quantization survives against doping. At a threshold doping, the topological domain ceases to percolate and the system is taken over by the $\nu=4/3$ domain, in which case the Hall conductance vanishes. One interesting feature of the experimental phase diagram of MoTe$_2$ moir\'{e} is that the Hall conductance plateau extends for
fillings around $\nu=1$ \citep{Park2023,PhysRevX.13.031037,park2024ferromagnetism,xu2024interplay}. The QAHC and DW phases offer new possible mechanisms to explain the robustness of quantization of the Hall conductance against doping, in addition to a more conventional mechanism based on the Anderson localization of doped electrons. In addition to Hall conductance, doping-induced spin textures can be detected experimentally through multiple techniques, including scanning tunneling microscope, electronic compressibility and optical measurements.

Interesting future directions include understanding whether quantum
fluctuations can melt the QAHC and give rise to exotic states of matter such as superconductivity. In fact,
we have shown that the QAHC can contain charge-2e skyrmions in some
regions of parameters. The condensation of these
charge-2e skyrmions would give rise to
superconductivity. Interestingly, this may also occur for the simpler
charge-e skyrmions. Since a skyrmion generates one flux quantum emergent magnetic field for electrons, the charge-e skyrmion composite object is a boson, in analogy to the composite boson due to the attachment of one flux quantum to an electron in the fractional quantum Hall systems.

\section{acknowledgments}
The authors would like to thank Di Xiao, Cristian Batista and Long Ju for fruitful discussions. The work is partially supported by the US DOE NNSA under Contract No. 89233218CNA000001 through the LDRD Program and was performed, in part, at the Center for Integrated Nanotechnologies, an Office of Science User Facility operated for the U.S. DOE Office of Science, under user proposals \#2018BU0010 and \#2018BU0083.

\section{Methods}

We employ the unrestricted self-consistent Hartree-Fock method in
real-space. The full mean-field Hamiltonian corresponding to Eq.$\,$\ref{eq:H} is given by (see Supplementary Information
for detailed derivation)

\begin{equation}
\begin{aligned} & H_{MF}=H_{0}+U\sum_{i,\sigma}\langle n_{i,-\sigma}\rangle n_{i,\sigma}-U\sum_{i,\sigma}\langle c_{i,-\sigma}^{\dagger}c_{i,\sigma}\rangle c_{i,\sigma}^{\dagger}c_{i,-\sigma}\\
 & +V\sum_{i,j,\sigma,\sigma'}\delta_{\langle i,j\rangle}\langle n_{j,\sigma'}\rangle n_{i\sigma}-V\sum_{i,j,\sigma,\sigma'}\delta_{\langle i,j\rangle}\langle c_{j,\sigma'}^{\dagger}c_{i,\sigma}\rangle c_{i\sigma}^{\dagger}c_{j\sigma'}\\
 & -U\sum_{i}(\langle n_{i,\up}\rangle\langle n_{i,\down}\rangle-\langle c_{i\up}^{\dagger}c_{i\down}\rangle\langle c_{i\down}^{\dagger}c_{i\up}\rangle)\\
 & -V\sum_{\langle i,j\rangle,\sigma,\sigma'}(\langle n_{i\sigma}\rangle\langle n_{j\sigma'}\rangle-\langle c_{j\sigma'}^{\dagger}c_{i\sigma}\rangle\langle c_{i\sigma}^{\dagger}c_{j\sigma'}\rangle),
\end{aligned}
\label{eq:HMF}
\end{equation}
where $H_{0}$ is the non-interacting part of the Hamiltonian in Eq.$\,$\ref{eq:H}
and $\langle\cdots\rangle$ denotes the average value with respect to the
mean-field ground state, to be determined self-consistently. The site-resolved
magnetization vector can be computed from the variational parameters
as ${\bf m}_{i}=(2\Re[\langle c_{i\up}^{\dagger}c_{i\down}\rangle],\ 2\Im[\langle c_{i\up}^{\dagger}c_{i\down}\rangle],\ \langle n_{i,\uparrow}\rangle-\langle n_{i,\downarrow}\rangle)$.
We considered finite system sizes with $L\times L$ unit cells. In
order to increase the speed of convergence we employ the Broyden method
\citep{PhysRevB.38.12807}. Furthermore, in order to avoid convergence
to local minima, we also performed between $100$ and $1000$ calculations
(depending on system size) with different starting guesses and took
the one with the smallest free energy as the ground state. Finally,
we use twisted boundary conditions with randomly chosen twists. This
allows to break unwanted degeneracies (or quasi-degeneracies) that
can be harmful for the convergence of the mean-field equations. 

For the smallest system sizes used ($L=12$ and $L=16$), we used
a completely random guess for the variational parameters in order
to identify the possible phases in an unbiased way. To provide the thermodynamic limit
estimation of the phase boundaries, we carried out calculations for
larger $L$ (up to $L=32$), verifying the convergence of the critical
points, within the provided error bars. Because for a large system
it becomes very challenging to converge from a random starting guess,
we used random skyrmion and domain wall initial guesses motivated
by the smaller $L$ results. We then compared the free energies and
converged variational parameters for both guesses. In the Supplementary
Information Sec.$\,$\ref{sec:SUP_phase_boundary_details} we provide
a detailed overview of these calculations. Finally, defining the vector
of variational parameters ${\bf v}=({\bf n}_{\up},{\bf n}_{\down},\Re[{\bf \Delta}_{\up\down}],\Im[{\bf \Delta}_{\up\down}],\Re[\tilde{{\bf t}}],\Im[\tilde{{\bf t}}])$,
where ${\bf \Delta}_{\up\down}^{i}=\langle c_{i,\up}^{\dagger}c_{i,\down}\rangle$
and $\tilde{{\bf t}}_{ij;\sigma\sigma'}=\langle c_{i\sigma}^{\dagger}c_{j\sigma'}\rangle\delta_{\langle i,j\rangle}$,
we only stopped the calculation when $|\Delta{\bf v}|<10^{-8}$, where
$\Delta{\bf v}$ denotes the difference between consecutive iterations,
up to a maximum of $N_{\textrm{iter}}$ iterations. Depending on the
size of the system, we chose $N_{\textrm{iter}}\in[10^{3}-5\times10^{4}]$.

\bibliography{refs}

%%%%%%%%%%%%%%%%%%%%%%%%%%%%%%%%%%%%%%%%%%%%%%%%%%%%%%%%%

%%%%%%%%%%%%%%%%%%%%%%%%%%%%%%%%%%%%%%%%%%%%%%%%%%%%%%%%%

%%%%%%%%%%%%%%%%%%%%%%%%%%%%%%%%%%%%%%%%%%%%%%%%%%%%%%%%%

\clearpage\onecolumngrid

\beginsupplement
\begin{center}
\textbf{\large{}Supplemental Information for: \vspace{0.1cm}
}{\large\par}
\par\end{center}
\title{Doping-induced Quantum Anomalous Hall Crystals and Topological Domain
Walls}

\vspace{0.3cm}

\tableofcontents{}

\newpage{}

\section{TMD homobilayers moir\'{e} superlattice as Kane-Mele-Hubbard model}

\begin{figure}[h!]
\centering{}\includegraphics[width=0.5\textwidth]{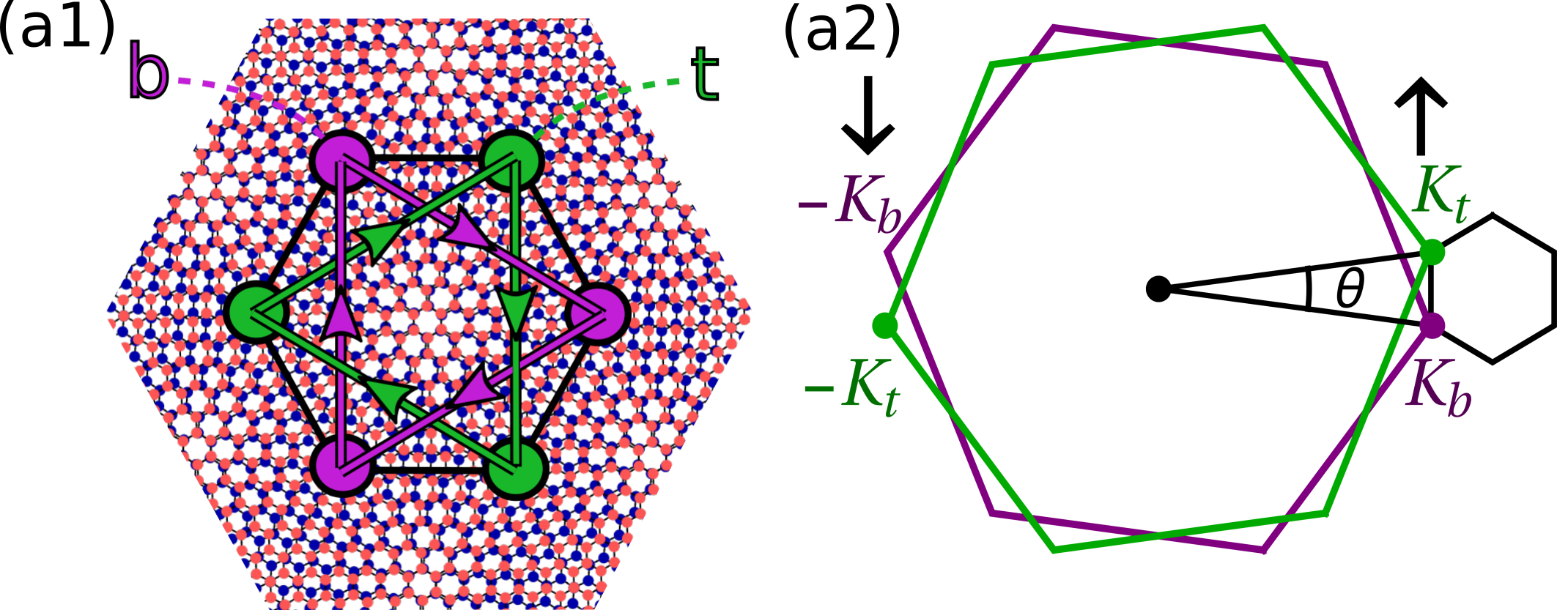}\caption{Illustration of emergent Kane-Mele-Hubbard model
in TMD homobilayer moiré superlattice. (a) shows a moir\'{e} cell and (b) shows the moir\'{e} mini Brillouin zone constructed from the twisted Brillouin zone of constituent layer. \label{fig:1a_SUP}}
\end{figure}

In the main text, we discuss that TMD homobilayers moir\'{e} superlattice can be described by an effective Kane-Mele-Hubbard model. Here we provide a more detailed explanation. 
The essential ingredients behind this emergent description are illustrated in Fig.$\,$\ref{fig:1}(a), that we reproduce here in Fig.$\,$\ref{fig:1a_SUP}. These include \citep{PhysRevLett.122.086402}: 
\begin{itemize}
    \item (i)
the Wannier functions for the two topmost valence bands of TMD homobilayers moir\'{e} superlattice, such as twisted MoTe$_2$ bilayer, are
very localized in the purple and green corners of the moiré cell depicted
in Fig.$\,$\ref{fig:1a_SUP}(a1), respectively for the bottom and
top layers;
    \item (ii) the maxima and minima of the first and second highest
energy valence bands are respectively located at the Dirac points
${\bf K}_{t}$ and ${\bf K}_{b}$ of the moiré Brillouin zone arising
from the displaced Dirac cones of top and bottom layers, as shown
in Fig.$\,$\ref{fig:1a_SUP}(a2);
    \item (iii) there is spin-valley locking,
that is, the wave function at ${\bf K}$ ($-{\bf K}$) valley is spin
up (down) polarized, as illustrated
in Fig.$\,$\ref{fig:1a_SUP}(a2).
\end{itemize}

Because of (i), a tight-binding model in a honeycomb
lattice with the emergent sites represented in Fig.$\,$\ref{fig:1a_SUP}(a1)
can be derived to describe the topmost valence bands. Also, because
of (ii), the low-energy expansion around ${\bf K}_{t}$ and ${\bf K}_{b}$
introduces layer(sublattice)-dependent next-nearest-neighbor (NNN)
complex hopping between the effective sites given by $\exp(\textrm{i}\sigma{\bf \kappa}_{t(b)}\cdot{\bf a}_{M})$,
where ${\bf a}_{M}$ is the moiré lattice vector connecting two NNN
sites and $\sigma=\pm1$ respectively for spins $\up$ and $\down$.
The spin-dependence of the phase follows from (iii). This NNN term
is precisely the Kane-Mele spin-orbit term, responsible for the non-trivial
topology. The bands with
opposite spin/valley have opposite Chern numbers. At twist angles when the electron kinetic energy is reduced, the interaction between electrons can be much larger than the
kinetic energy. The most important interaction term is on-site Hubbard,
even though longer-range interactions can also be large \citep{doi:10.1073/pnas.2316749121}.

\section{Easy-axis ferromagnetic QAHI for $\nu=1$}

\label{sec:SUP_QAHI}

In this section we obtain the homogeneous ferromagnetic QAHI state
at filling $\nu=1$. To keep the problem analytically tractable,
we will first search for homogeneous solutions. We will then confirm
indeed that the true ground state is homogeneous by carrying out the full
real-space mean-field calculation for this filling. 

Let us assume that

\begin{equation}
\begin{array}{c}
\langle n_{i,\sigma}\rangle=\mathcal{N}_{\sigma}\equiv\frac{1}{N_{{\rm uc}}}\sum_{i}\langle n_{i,-\sigma}\rangle\\
\langle c_{i,-\sigma}^{\dagger}c_{i,\sigma}\rangle=\Delta_{\sigma,-\sigma}\equiv\frac{1}{N_{{\rm uc}}}\sum_{i}\langle c_{i,-\sigma}^{\dagger}c_{i,\sigma}\rangle
\end{array}
\end{equation}
where $N_{\textrm{uc}}$ is the number of unit cells. Using this assumption,
we get the following mean-field Hamiltonian

\begin{equation}
\begin{array}{cc}
H_{MF} & =H_{0}+U\Big(\sum_{i,\sigma}\mathcal{N}_{-\sigma}n_{i,\sigma}-\sum_{i,\sigma}\Delta_{\sigma,-\sigma}c_{i,\sigma}^{\dagger}c_{i,-\sigma}\Big)\end{array},\label{eq:MF_homogeneous}
\end{equation}
where by hermiticity $\Delta_{\up,\down}=(\Delta_{\down,\up})^{*}$.
Introducing the Fourier transforms 

\begin{equation}
c_{i,\sigma}^{\dagger}=\begin{cases}
\frac{1}{\sqrt{N_{{\rm uc}}}}\sum_{{\bf k}}e^{{\rm i}{\bf k}\cdot{\bf R}_{i}}a_{{\bf k},\sigma}^{\dagger} & ,i\in A\\
\frac{1}{\sqrt{N_{{\rm uc}}}}\sum_{{\bf k}}e^{{\rm i}{\bf k}\cdot{\bf R}_{i}}b_{{\bf k},\sigma}^{\dagger} & ,i\in B
\end{cases}
\end{equation}
the momentum-space mean-field Hamiltonian reads 
\begin{equation}
H=\sum_{{\bf k}}{\bf c}_{{\bf k}}^{\dagger}{\bf h}_{{\bf k}}^{{\rm MF}}{\bf c}_{{\bf k}}+\textrm{cte.}\,,\label{eq:k_space_MF}
\end{equation}
where $\textrm{cte}$ is a constant with no fermionic operators and

${\bf c}_{{\bf k}}=(\begin{array}{cccc}
a_{{\bf k},\up} & b_{{\bf k},\up} & a_{{\bf k},\down} & b_{{\bf k},\down}\end{array})^{T}$

\begin{equation}
{\bf h}_{{\bf k}}^{{\rm MF}}=\left(\begin{array}{cccc}
\lambda({\bf k})+U\mathcal{N}_{\down} & f({\bf k}) & -U\Delta_{\down\up} & 0\\
f^{*}({\bf k}) & -\lambda({\bf k})+U\mathcal{N}_{\down} & 0 & -U\Delta_{\down\up}\\
-U\Delta_{\down\up}^{*} & 0 & -\lambda({\bf k})+U\mathcal{N}_{\up} & f({\bf k})\\
0 & -U\Delta_{\down\up}^{*} & f^{*}({\bf k}) & \lambda({\bf k})+U\mathcal{N}_{\up}
\end{array}\right)
\end{equation}

\begin{equation}
f({\bf k})=-t(1+e^{{\rm i}{\bf k}\cdot{\bf a}_{1}}+e^{{\rm i}{\bf k}\cdot{\bf a}_{2}})
\end{equation}

\begin{equation}
\lambda({\bf k})=-2t_{2}[\sin({\bf k}\cdot{\bf a}_{1})-\sin({\bf k}\cdot({\bf a}_{1}-{\bf a}_{2}))-\sin({\bf k}\cdot{\bf a}_{2})]
\end{equation}

Defining

\begin{equation}
\begin{cases}
m_{x}=\Delta_{\down\up}+\Delta_{\up\down}\\
m_{y}=-i(\Delta_{\up\down}-\Delta_{\down\up})\\
m_{z}=\mathcal{N}_{\up}-\mathcal{N}_{\down}
\end{cases}
\end{equation}
we can write the mean-field Hamiltonian as

\begin{equation}
H_{MF}=\sum_{{\bf k}}{\bf c}_{{\bf k}}^{\dagger}\left[\frac{U\mathcal{N}}{2}\mathbb{I}+f_{x}({\bf k})\sigma_{x}+f_{y}({\bf k})\sigma_{y}+f_{z}({\bf k})\sigma_{z}s_{z}-\frac{U}{2}{\bf m}\cdot{\bf s}\right]{\bf c}_{{\bf k}}+\frac{U}{2}N_{\textrm{uc}}{\bf m}^{2}-\frac{U}{2}N_{\textrm{uc}}n^{2}
\end{equation}

where $\mathcal{N}=\mathcal{N}_{\up}+\mathcal{N}_{\down}$, ${\bf m}=(m_{x},m_{y},m_{z})$,
${\bf s}$ are the Pauli matrices in the spin space and we made explicit
the constant term in Eq.$\,$\ref{eq:k_space_MF} since it depends
on ${\bf m}$. Writing ${\bf m}=M(\sin\theta\cos\phi,\sin\theta\sin\phi,\cos\theta)$,
the eigenenergies are given by

\begin{equation}
\epsilon({\bf k})=f_{0}({\bf k})+\frac{U\mathcal{N}}{2}\pm\sqrt{{\bf f}^{2}({\bf k})+(UM/2)^{2}\pm|UM|\sqrt{f_{x}^{2}({\bf k})+f_{y}^{2}({\bf k})+f_{z}^{2}({\bf k})\cos^{2}(\theta)}}
\end{equation}

Let us consider the situation where $U$ is large enough so that the
lowest energy band is isolated ($U\gtrsim4$ for $t_{2}=0.2$). In
this case, we simply need to consider the free energy of the lowest
energy band, which is minimized for $\theta=0$. Therefore, the ground state
develops finite magnetization along the $z$ axis. Computing the total
free energy for $\theta=0$, we have

\begin{equation}
F=\langle H_{MF}\rangle=\langle\sum_{\alpha,{\bf k}}\epsilon_{\alpha}({\bf k},M)c_{{\bf k},\alpha}^{\dagger}c_{{\bf k},\alpha}\rangle+UN_{\textrm{uc}}(M^{2}-n^{2})=\sum_{{\bf k}}\Big(f_{0}({\bf k})-|{\bf f}^{2}({\bf k})|\Big)+\frac{UN_{\textrm{uc}}}{2}(-|M|+M^{2})+\textrm{cte},\label{eq:free_energy_1o4}
\end{equation}
which is minimized for $M=1/2$, consistent with the numerical results.

\section{Derivation of mean-field Hamiltonian}

\label{sec:MF_derivation}

Starting with the Hamiltonian in Eq.$\,$\ref{eq:H} of the main text,
we can make the following mean-field decoupling by using Wick's theorem:

\begin{equation}
\begin{array}{cc}
n_{i,\sigma}n_{j,\sigma'}=c_{i\sigma}^{\dagger}c_{j\sigma'}^{\dagger}c_{j\sigma'}c_{i\sigma}=-:c_{i\sigma}^{\dagger}c_{j\sigma'}^{\dagger}c_{j\sigma'}c_{i\sigma}:\\
=\langle n_{i\sigma}\rangle:n_{j\sigma'}:+\langle n_{j\sigma'}\rangle:n_{i\sigma}:+\langle n_{i\sigma}\rangle\langle n_{j\sigma'}\rangle\\
-\langle c_{i\sigma}^{\dagger}c_{j\sigma'}\rangle:c_{j\sigma'}^{\dagger}c_{i\sigma}:-\langle c_{j\sigma'}^{\dagger}c_{i\sigma}\rangle:c_{i\sigma}^{\dagger}c_{j\sigma'}:-\langle c_{j\sigma'}^{\dagger}c_{i\sigma}\rangle\langle c_{i\sigma}^{\dagger}c_{j\sigma'}\rangle\\
\textrm{Using \ensuremath{:c_{i\sigma}^{\dagger}c_{j\sigma'}:=c_{i\sigma}^{\dagger}c_{j\sigma'}-\langle c_{i\sigma}^{\dagger}c_{j\sigma'}\rangle} }\\
=\langle n_{i\sigma}\rangle n_{j\sigma'}+\langle n_{j\sigma'}\rangle n_{i\sigma}-\langle n_{i\sigma}\rangle\langle n_{j\sigma'}\rangle\\
-\langle c_{i\sigma}^{\dagger}c_{j\sigma'}\rangle c_{j\sigma'}^{\dagger}c_{i\sigma}-\langle c_{j\sigma'}^{\dagger}c_{i\sigma}\rangle c_{i\sigma}^{\dagger}c_{j\sigma'}+\langle c_{j\sigma'}^{\dagger}c_{i\sigma}\rangle\langle c_{i\sigma}^{\dagger}c_{j\sigma'}\rangle
\end{array}
\end{equation}
where $::$ denotes the normal ordering operation. Plugging this in
the Hamiltonian in Eq.$\,$\ref{eq:H}, we get the meanfield Hamiltonian
in Eq.$\,$\ref{eq:HMF}. 

The expectation values can be computed at every mean-field iteration
by introducing the eigenbasis $d_{\alpha}=\sum_{i,\sigma}\mathcal{U}_{i\sigma,\alpha}^{*}c_{i,\sigma}$,
through 

\begin{equation}
\begin{array}{cc}
\langle c_{j\sigma}^{\dagger}c_{i\sigma'}\rangle & =\Big\langle\sum_{\alpha,\beta}d_{\alpha}^{\dagger}\mathcal{U}_{j\sigma,\alpha}^{*}\mathcal{U}_{i\sigma',\beta}d_{\beta}\Big\rangle_{MF}\\
 & =\sum_{\alpha}\mathcal{U}_{j\sigma,\alpha}^{*}\mathcal{U}_{i\sigma',\alpha}f(E_{\alpha})
\end{array}
\end{equation}
where $f(E_{\alpha})=[\exp[\beta(E_{\alpha}-\mu)]+1]^{-1}$. Finally,
the free energy can be computed through 

\begin{equation}
F=\langle H\rangle=\langle H_{0}\rangle+\langle H_{U}\rangle+\langle H_{V}\rangle\,,
\end{equation}
where

\begin{equation}
\begin{array}{cc}
\langle H_{0}\rangle_{MF}=-\sum_{i,\sigma}\langle t\sum_{\langle i,j\rangle}c_{i,\sigma}^{\dagger}c_{j,\sigma}-t_{2}\sum_{\langle\langle i,j\rangle\rangle,\sigma}e^{-{\rm i}\sigma\phi_{ij}}c_{i,\sigma}^{\dagger}c_{j,\sigma}\rangle_{MF}\\
=\sum_{\alpha\in{\rm occ}}\Big(-t\sum_{\langle i,j\rangle}\mathcal{U}_{\alpha,i\sigma}^{*}\mathcal{U}_{\alpha,j\sigma}-t_{2}\sum_{\langle\langle i,j\rangle\rangle,\sigma}e^{-{\rm i}\sigma\phi_{ij}}\mathcal{U}_{\alpha,i\sigma}^{*}\mathcal{U}_{\alpha,j\sigma}\Big)\\
=\Tr_{{\rm occ}}[\mathcal{U}^{\dagger}H_{0}\mathcal{U}]
\end{array}
\end{equation}

\begin{equation}
\begin{array}{cc}
\langle H_{U}\rangle_{MF}= & \langle U\sum_{i}n_{i,\up}n_{i,\down}\rangle_{MF}=U\sum_{i}\langle n_{i,\up}\rangle_{MF}\langle n_{i,\down}\rangle_{MF}-U\langle c_{i\up}^{\dagger}c_{i\down}\rangle_{MF}\langle c_{i\down}^{\dagger}c_{i\up}\rangle_{MF}\\
 & =U\sum_{i}(\sum_{\alpha\in{\rm occ}}|\mathcal{U}_{i\up,\alpha}|^{2})(\sum_{\beta\in{\rm occ}}|\mathcal{U}_{i\down,\beta}|^{2})-U\sum_{i}|\sum_{\alpha\in{\rm occ}}\mathcal{U}_{i\up,\alpha}\mathcal{U}_{i\down,\alpha}^{*}|^{2}
\end{array}
\end{equation}

\begin{equation}
\begin{array}{cc}
\langle H_{V}\rangle_{MF}= & \langle V\sum_{\langle i,j\rangle,\sigma,\sigma'}n_{i\sigma}n_{j\sigma'}\rangle=V\sum_{\langle i,j\rangle,\sigma,\sigma'}\langle n_{i\sigma}\rangle_{MF}\langle n_{j\sigma'}\rangle_{MF}-V\langle c_{i\sigma}^{\dagger}c_{j\sigma'}\rangle_{MF}\langle c_{j\sigma'}^{\dagger}c_{i\sigma}\rangle_{MF}\\
 & =V\sum_{\langle i,j\rangle,\sigma,\sigma'}(\sum_{\alpha\in{\rm occ}}|\mathcal{U}_{i\sigma,\alpha}|^{2})(\sum_{\beta\in{\rm occ}}|\mathcal{U}_{j\sigma',\beta}|^{2})-V\sum_{\langle i,j\rangle,\sigma,\sigma'}|\sum_{\alpha\in{\rm occ}}\mathcal{U}_{i\sigma,\alpha}^{*}\mathcal{U}_{j\sigma',\alpha}|^{2}
\end{array}
\end{equation}

\section{Details on calculation of phase boundaries}

\label{sec:SUP_phase_boundary_details}

\begin{figure}[H]
\begin{centering}
\includegraphics[width=0.85\columnwidth]{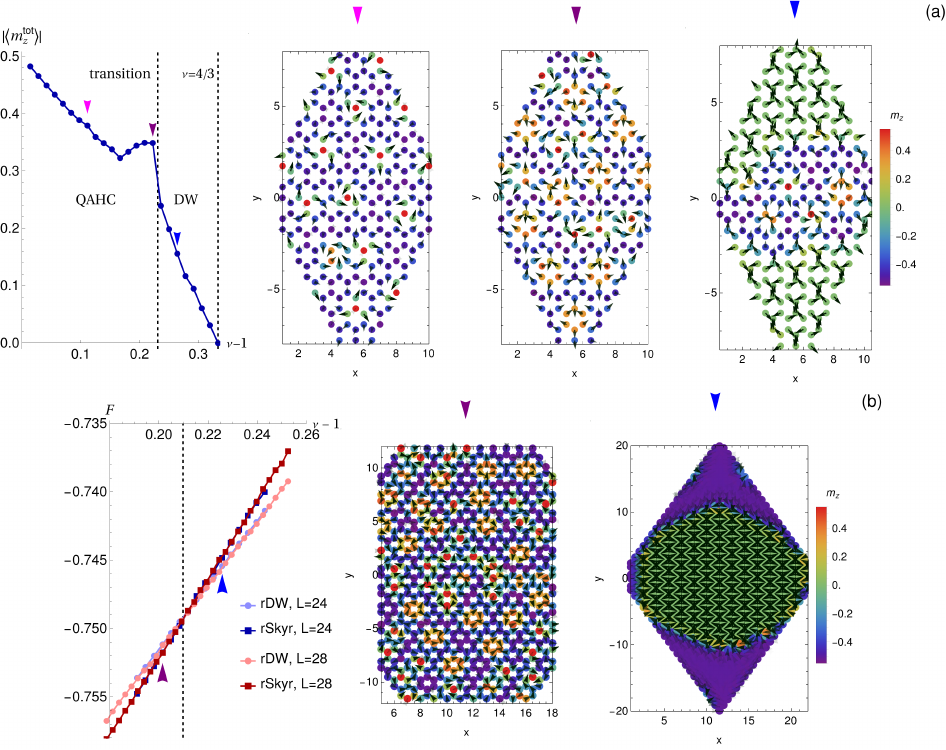}
\par\end{centering}
\caption{Results for the QAHC to DW transition, for $t_{2}=0.075$,
$U=8$. Left: $|\langle m_{z}^{\textrm{tot}}\rangle|=N_{\textrm{sites}}^{-1}|\sum_{i}\langle m_{i}^{z}\rangle|$
as a function of $(\nu-1)$ for $L=12$ and a fully random starting
guess. The dashed line separates the QAHC and DW phases, signaled
by a discontinuity in $|\langle m_{z}^{\textrm{tot}}\rangle|$. Right:
The magnetization profiles for the fillings indicated
by the arrows. (b) Left: Free energies for initial rDW and rSkyr starting
guesses, for sizes $L=24,28$. The critical point is estimated through
the crossing point. Right. Magnetization profiles for the
rSkyr (purple arrow) and rDW (blue arrow) starting guesses, for $L=24$.\label{fig:SUP_QAHC_to_DW}}
\end{figure}

In this section we provide details on the calculations of the phase
boundaries presented in the main text. As mentioned in the \textit{Methods}
section, we started by running completely unbiased calculations with
a random starting guess for smaller system sizes, while for the larger
systems we provided educated guesses. We show an example of this scheme
in Fig.$\,$\ref{fig:SUP_QAHC_to_DW}. In Fig.$\,$\ref{fig:SUP_QAHC_to_DW}(a)
we show the results for $L=12$ and a completely random starting guess,
where the transition between the QAHC and the DW phases can be clearly
observed. In order to determine the critical point as precisely as
possible, we provided the following educated starting guesses for
the larger system sizes:
\begin{itemize}
\item Random domain wall (rDW): circular coplanar magnetic domain with
random radius (around a mean value determined by number of doped electrons)
in a ferromagnetic background;
\item Random skyrmions (rSkyr): $\delta_{e}$ skyrmions with $\chi=1$ charge,
randomly distributed in the lattice over a ferromagnetic background.
\end{itemize}
Given the starting guess rDW (rSkyr), even if the true ground state
is the QAHC (DW), the system may converge to a DW (QAHC) state corresponding
to a local minimum in the free energy. Nonetheless, the true ground state
can be unveiled by comparing the free energies of the two cases, which
show a crossing point that is robust to increasing $L$, as exemplified
in Fig.$\,$\ref{fig:SUP_QAHC_to_DW}(b). This crossing point provides
an accurate estimation of the critical point between the QAHC and
DW phases, whose convergence can be tested by comparing the results
for different system sizes.

\begin{figure}[H]
\begin{centering}
\includegraphics[width=1\columnwidth]{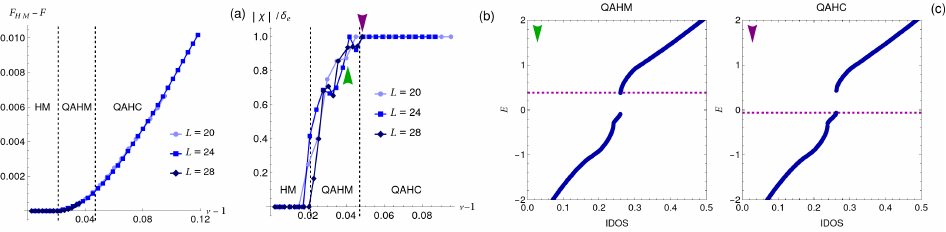}
\par\end{centering}
\caption{Results for the transitions between the HM, QAHM and QAHC
phases, for $t_{2}=0.05$ and $U=8$. (a) Difference between the ground state
and HM free energies. For the ground state we used the rSkyr starting
guess. (b) $|\chi|/\delta_{e}$ for different $L$. (c) Eigenenergies
as a function of integrated density of states (IDOS) for the dopings
indicated by the arrows in (b), for $L=28$. The purple dashed line
indicates the Fermi level. \label{fig:SUP_HM_to_QAHM_to_QAHC} }
\end{figure}

We now turn to explain how the phase boundaries of the HM and QAHM
phases were obtained in Fig.$\,$\ref{fig:1}(c). In Fig.$\,$\ref{fig:SUP_HM_to_QAHM_to_QAHC}(a)
we show the difference in free energy between the true ground state
and the homogeneous HM state. Motivated by the results for smaller systems
that showed that the ground state is a QAHC for this region of parameters,
we used rSkyr as a starting guess for larger system sizes. For
low enough doping, the free energies of the HM and the true ground state
are exactly the same, since there is convergence to the HM state even
with the inhomogeneous rSkyr starting guess. Above a critical doping,
this difference starts being finite, and the true ground state becomes
the QAHM. This critical doping is robust to increasing $L$ as shown
in Fig.$\,$\ref{fig:SUP_HM_to_QAHM_to_QAHC}(a). The QAHM phase is
characterized by a skyrmion charge $0<\chi<\delta_{e}$ {[}Fig.$\,$\ref{fig:SUP_HM_to_QAHM_to_QAHC}(b){]},
a non-quantized Hall conductance {[}inset of Fig.$\,$\ref{fig:1}(c){]}
and a gapless spectrum {[}Fig.$\,$\ref{fig:SUP_HM_to_QAHM_to_QAHC}(c){]}.
Above a critical doping there is a transition to QAHC, characterized
by $\chi=\delta_{e}$ {[}Fig.$\,$\ref{fig:SUP_HM_to_QAHM_to_QAHC}(b){]},
a robust gap {[}Fig.$\,$\ref{fig:SUP_HM_to_QAHM_to_QAHC}(c){]} and
$|\sigma_{xy}|=e^{2}/h$ {[}inset of Fig.$\,$\ref{fig:1}(c){]}.
The critical point was estimated by averaging the critical points
obtained for the two largest system sizes.

\begin{figure}[H]
\begin{centering}
\includegraphics[width=0.6\columnwidth]{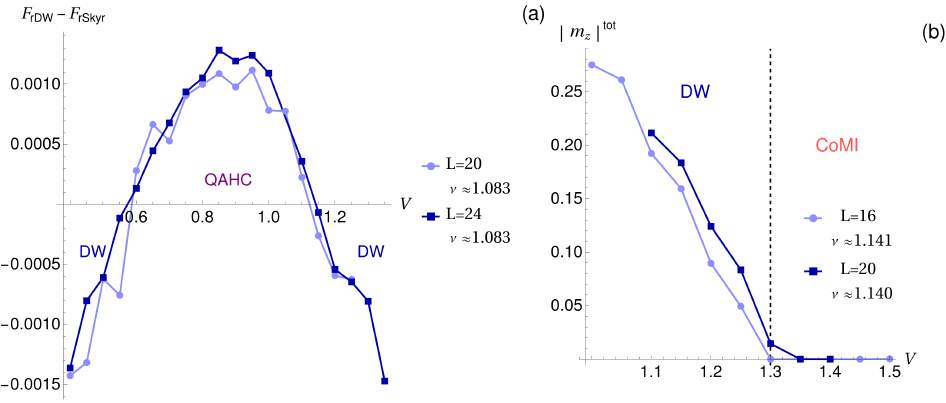}
\par\end{centering}
\caption{Results for the transitions in Fig.$\,$\ref{fig:5}(a), with
finite nearest-neighbor interactions, for $t_{2}=0.2$ and $U=8$.
(a) Difference between the free energies for the mean-field solution
obtained using rDW and rSkyr starting guesses, for $\nu\approx1.084$.
(b) $|m_{z}|^{\textrm{tot}} = N_{\textrm{sites}}^{-1}\sum_{i} |\langle m_{i}^{z}\rangle|$ for $\nu\approx1.14$,
for a fully random starting guess. \label{fig:SUP_NN_transitions}}
\end{figure}

We finally detail the calculation of the phase boundaries for the
phase diagram in Fig.$\,$\ref{fig:5}(a), with finite nearest-neighbor
interactions. For the DW to QAHC transition we used the same procedure
as the one previously described, showing an example in Fig.$\,$\ref{fig:SUP_NN_transitions}(a).
For the transition between the DW and CoMI phases, we inspected the
critical $V$ above which $|m_{z}|^{\textrm{tot}} = N_{\textrm{sites}}^{-1}\sum_{i} |\langle m_{i}^{z}\rangle|$ vanishes,
as exemplified in Fig.$\,$\ref{fig:SUP_NN_transitions}(b). Note that $|m_{z}|^{\textrm{tot}}$ decreases smoothly with $V$, corroborating the second-order nature of the transition, where the correlation length associated with the coplanar domain diverges. In this
case, we used a completely random initial guess and therefore reached
smaller system sizes (up to $L=20$) than for the DW to QAHC transition.
Nonetheless, a reasonable convergence in the critical point was obtained
for the two largest system sizes ($L=16$ and $L=20$), as indicated
by the small error bars in Fig.$\,$\ref{fig:5}(a), that correspond
to the difference in the critical point estimations for these sizes.

\section{A deeper look into the different phases}

\label{sec:SUP_different_phases}

In this section we explore in more detail some interesting regions
of the phase diagrams shown in the main text. 

\paragraph*{Electron densities in DW and QAHC phases.---}

\begin{figure}[H]

\begin{centering}
\includegraphics[width=0.9\columnwidth]{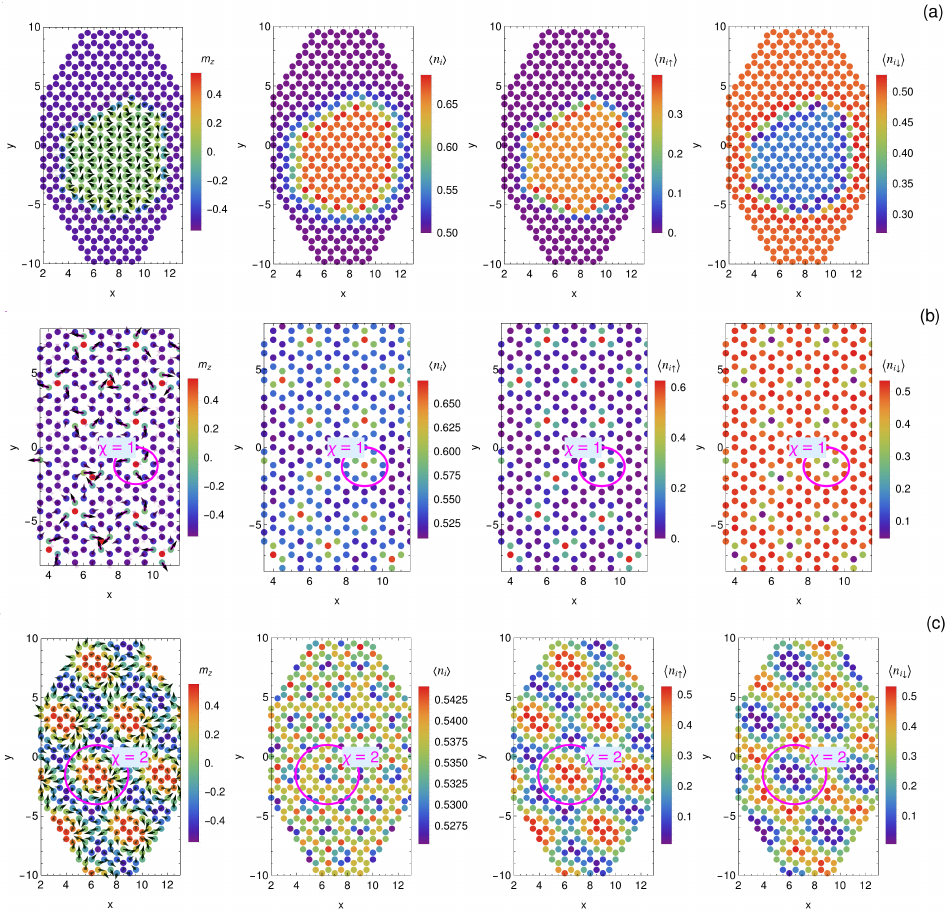}\caption{Magnetization and density profiles for the examples shown in Fig.$\,$\ref{fig:1}(d)
of the main text. Results for $L=16$ and (a) $t_{2}=0.125,\delta_{e}=26$
(DW); (b) $t_{2}=0.1,\delta_{e}=26$ (QAHC); and (c) $t_{2}=0,\delta_{e}=18$
(QHAC).\label{fig:Magnetization-and-density}}
\par\end{centering}
\end{figure}

In Fig.$\,$\ref{fig:1}(d) of main text we presented examples of
the magnetization profiles for the DW and QAHC phases. In Fig.$\,$\ref{fig:Magnetization-and-density}
we present the electron densities $\langle n_{i}\rangle=\langle n_{i\uparrow}\rangle+\langle n_{i\downarrow}\rangle$,
$\langle n_{i\uparrow}\rangle$ and $\langle n_{i\downarrow}\rangle$
together with the magnetization profiles. Within the DW phase, Fig.$\,$\ref{fig:Magnetization-and-density}(a)
shows that the density profile within the coplanar domain corresponds
to $\langle n_{i\uparrow}\rangle=\langle n_{i\downarrow}\rangle=1/3$,
while in the ferromagnetic domain we have $\langle n_{i\uparrow}\rangle=0$
and $\langle n_{i\downarrow}\rangle=1/2$ (note that we can also converge
to the alternative degenerate symmetry breaking state with $\langle n_{i\uparrow}\rangle=1/2$
and $\langle n_{i\downarrow}\rangle=0$). In the QAHC phase with charge-e
and $\chi=1$ skyrmions {[}Fig.$\,$\ref{fig:Magnetization-and-density}(b){]},
the plot of $\langle n_{i}\rangle$ shows a charge density wave modulation,
with the excess charge accumulating around the skyrmions. The excess
charge around each skyrmion compared to the ferromagnetic background
charge $\langle n_{i}\rangle=0.5$ is precisely one electron, as mentioned
in the main text. Finally, in the QAHC phase with charge-2e and $\chi=2$
skyrmions {[}Fig.$\,$\ref{fig:Magnetization-and-density}(c){]},
the $\langle n_{i}\rangle$ plot does not distinguish very well the
skyrmions from the background since they become significantly larger.
This becomes more clear in the spin-resolved plots, where spin $\up$
density is mostly distributed within the skyrmions while spin $\down$
density mostly occupies the (small) background.

\paragraph*{Ground states for small $\delta_{e}$.---}

\begin{figure}[H]
\begin{centering}
\includegraphics[width=0.85\columnwidth]{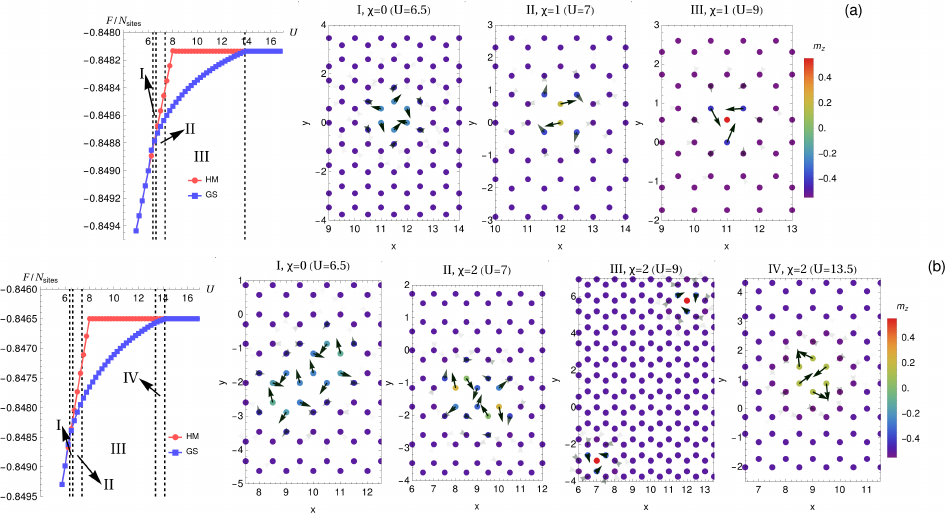}
\par\end{centering}
\caption{Ground states with different skyrmion charges for dopings $\delta_{e}=1,2$,
with $t_{2}=0.2$. Results for $\delta_{e}=1$, $L=24$ (a) and $\delta_{e}=2$,
$L=20$ (b), for the parameters indicated in the figure. The regions
with similar ground states as the ones represented in the rightmost
figures are depicted in the leftmost figures, and separated by black
dashed lines. \label{fig:SUP_delta-1.delta-2}}
\end{figure}

\begin{figure}[H]
\begin{centering}
\includegraphics[width=0.85\columnwidth]{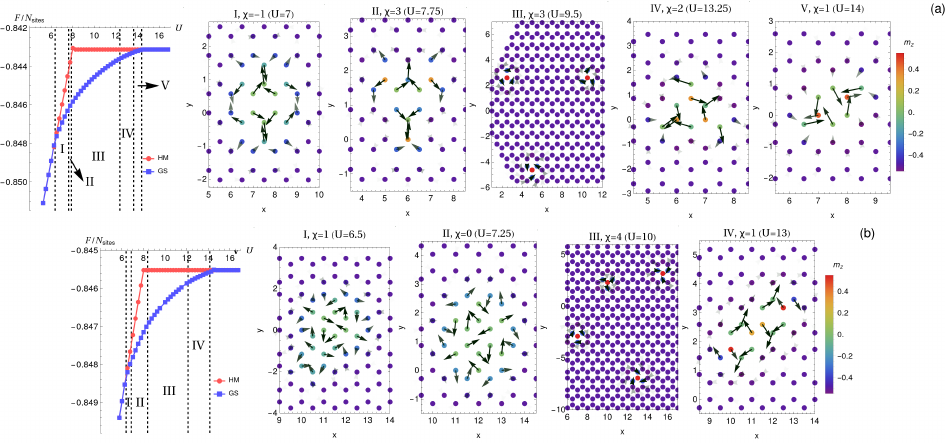}
\par\end{centering}
\caption{Ground states with different skyrmion charges for dopings $\delta_{e}=3,4$.
Results for $\delta_{e}=3$, $L=16$ (a) and $\delta_{e}=4$, $L=24$
(b), for the parameters indicated in the figure. \label{fig:SUP_delta-3.delta-4}}
\end{figure}

We now turn to analyze the phase diagram in Fig.$\,$\ref{fig:2}(a)
in more detail, where doping of a small number of electrons was
considered with respect to filling $\nu=1$. While we decided to
label any region where a charge clustering was observed as a DW phase,
these regions have rich substructures. We unveil them in Figs.$\,$\ref{fig:SUP_delta-1.delta-2},\ref{fig:SUP_delta-3.delta-4}
for $1\leq\delta_{e}\leq4$. Interestingly, ground states with finite
skyrmion charge $\chi\leq\delta_{e}$ can be found in the DW region.
We note that for $\delta_{e}=1$ we considered the ground state in
region I, distinct from the skyrmion states in regions II and III
{[}see Fig.$\,$\ref{fig:SUP_delta-1.delta-2}(a){]}, to belong to
the DW phase, even though it does not correspond to any charge clustering
with the single-electron doping.

\paragraph*{DW vs. $\textrm{DW}_{2}$ phases.---}

\begin{figure}[H]
\begin{centering}
\includegraphics[width=0.8\columnwidth]{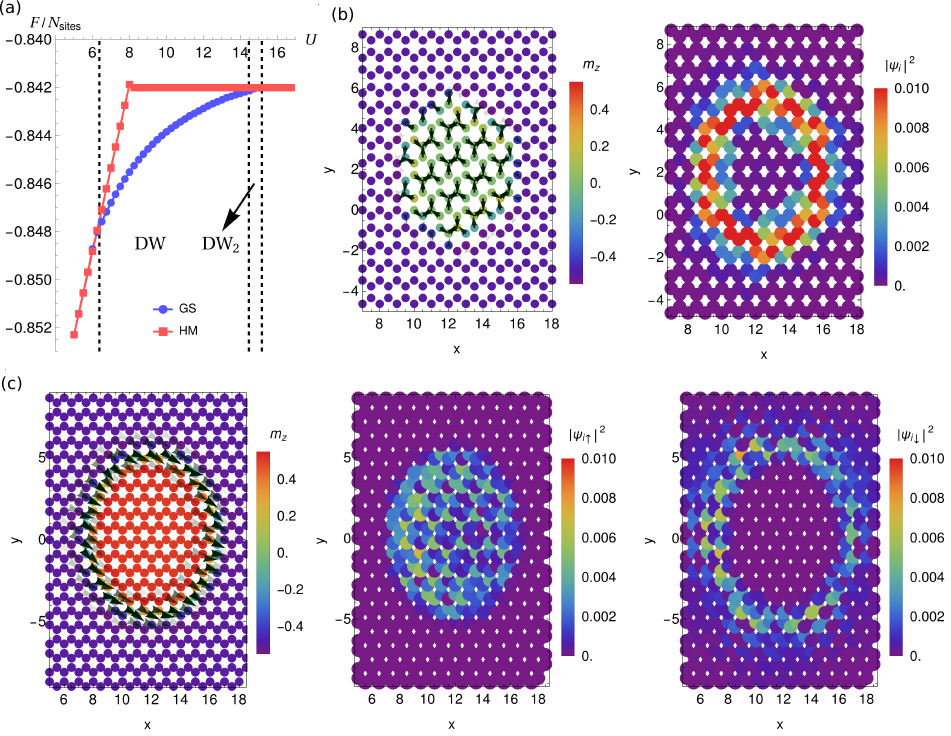}
\par\end{centering}
\caption{Ground states for $\delta_{e}=8$, with $t_{2}=0.2$. (a) Free energy
calculations for $L=24$. (b) Magnetization profile (left)
and spatial profile of the eigenstate at the Fermi level  (right, with $|\psi_{i}|^{2}=|\psi_{i\uparrow}|^{2}+|\psi_{i\downarrow}|^{2}$)
for $U=10$, within the DW phase. (c) Same as (b), for $U=14.5$ within
the $\textrm{DW}_{2}$ phase. \label{fig:SUP_delta-8_DW_DW2}}
\end{figure}

For larger electron dopings, we enter the regime where only the DW
phase survives {[}see examples in Figs.$\,$\ref{fig:SUP_delta-8_DW_DW2}(a,b)
for $\delta=8${]}. Over a small range of interaction strength, a
second domain wall phase that we label $\textrm{DW}_{2}$ is stabilized.
The magnetization profile in this phase is exemplified in Fig.$\,$\ref{fig:SUP_delta-8_DW_DW2}(c),
showing that two ferromagnetic domains with opposite spin polarization
are formed. However, in this case there are no clear chiral edge states
at the domain wall, which we attribute to the smaller domain being
metallic. This is evidenced by the extended nature of the eigenstate at the Fermi level within this domain, shown in Fig.$\,$\ref{fig:SUP_delta-8_DW_DW2}(c).
If this domain was insulating, then chiral edge states would be expected
at the domain wall due to the different Chern numbers for opposite
spin domains.

\paragraph*{A myriad of different skyrmion crystals.---}

\begin{figure}[H]
\centering{}\includegraphics[width=0.7\columnwidth]{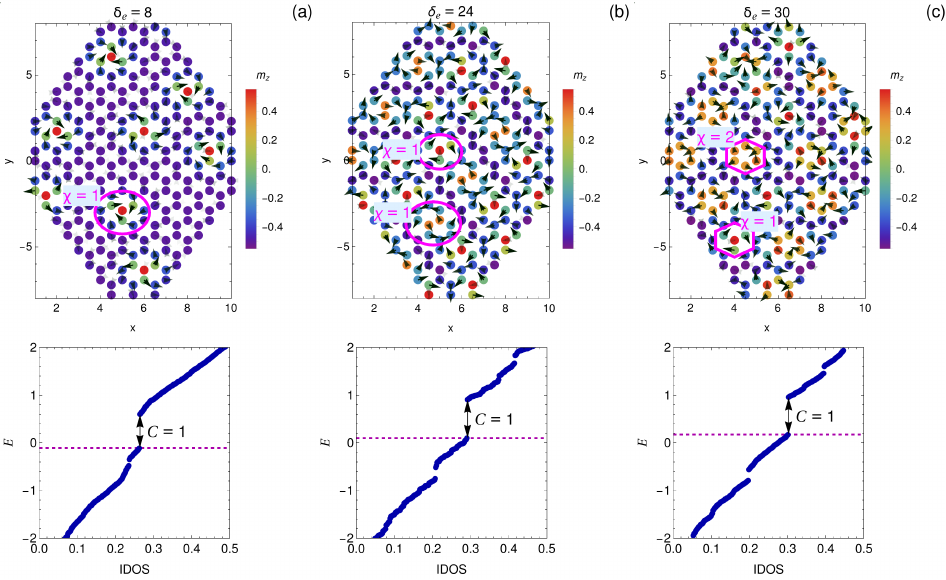}\caption{Different types of skyrmion crystals induced by doping, for $t_{2}=0.075,U=8$
and $L=12$ (using a fully random starting guess). Magnetization profiles
(top) together with IDOS plots (bottom). The purple dashed line indicates
the Fermi level. Results for (a) $\delta_{e}=8$ ($\nu\approx1.056$);
(b) $\delta_{e}=24$ ($\nu\approx1.168$) and (c) $\delta_{e}=24$
($\nu\approx1.208$). \label{fig:SUP_t2-0.075_doping}}
\end{figure}

\begin{figure}[H]
\centering{}\includegraphics[width=0.7\columnwidth]{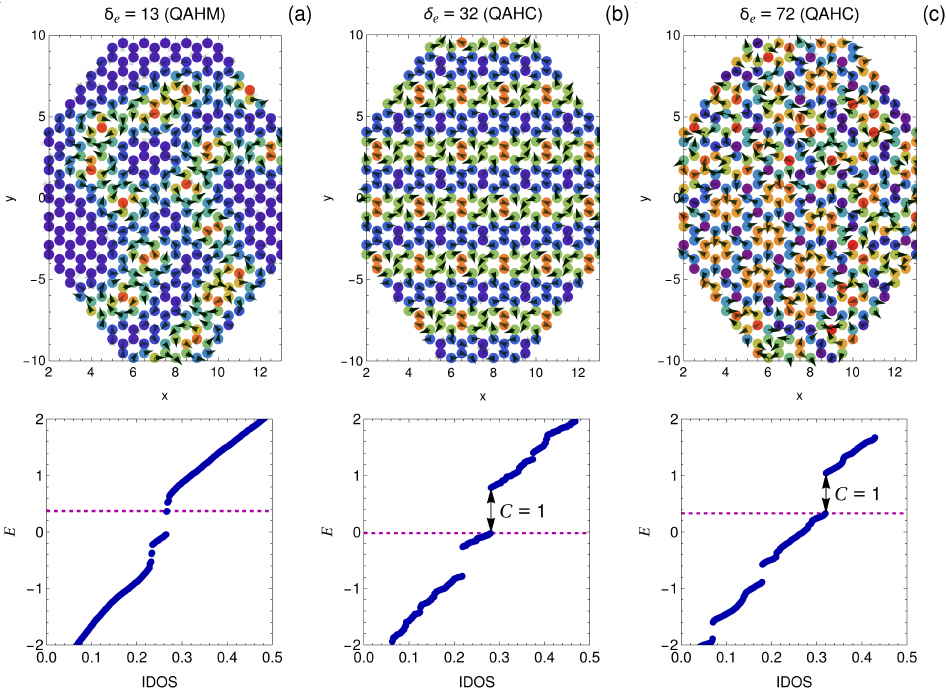}\caption{Different types of magnetic textures induced by doping, for $t_{2}=0.025,U=8$
and $L=16$ (using a fully random starting guess). Results for (a)
$\delta_{e}=13$ ($\nu\approx1.052$, in the QAHM phase); (b)
$\delta_{e}=40$ ($\nu\approx1.156$, in the QAHC phase) and (c)
$\delta_{e}=62$ ($\nu\approx1.24$, in the QAHC phase).\label{fig:SUP_t2-0.025_doping}}
\end{figure}

We now turn to explore the different types of skyrmion crystals that
can be stabilized in the QAHC phase by varying doping and the model
parameters. In Fig.$\,$\ref{fig:SUP_t2-0.075_doping} we show how
doping can induce charge-2e skyrmions with $\chi=2$ {[}see also Fig.$\,$\ref{fig:SUP_t2-0.025_doping}(c){]}.
The reason is that increasing doping decreases the lattice spacing
of the skyrmion crystal and given that the individual skyrmions repel,
it becomes energetically favorable to create slightly larger skyrmions
that accomodate more than one electron. 

\begin{figure}[H]
\centering{}\includegraphics[width=0.7\columnwidth]{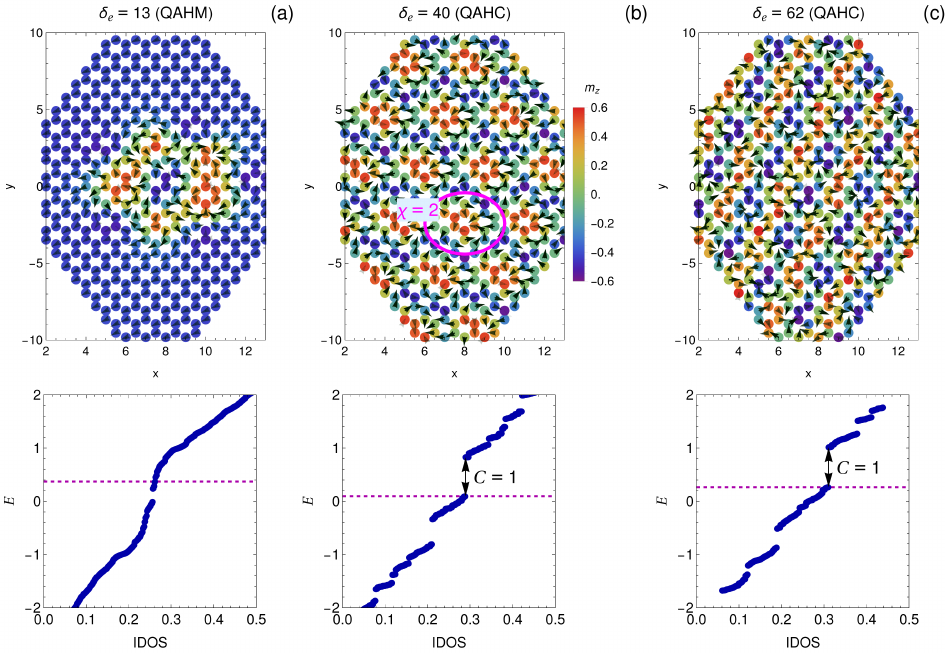}\caption{Different types of magnetic textures induced by doping, for $t_{2}=0,U=8$
and $L=16$ (using a fully random starting guess). Results for (a)
$\delta_{e}=18$ ($\nu\approx1.072$, in the QAHM phase); (b)
$\delta_{e}=32$ ($\nu\approx1.124$, in the QAHC phase) and (c)
$\delta_{e}=72$ ($\nu\approx1.28$, in the QAHC phase).\label{fig:SUP_t2-0_doping}}
\end{figure}

For smaller values of the topological mass $t_{2}$, the QAHM phase
can be stabilized, where skyrmions are formed but do not crystallize,
as exemplified in Figs.$\,$\ref{fig:SUP_t2-0.025_doping}(a),\ref{fig:SUP_t2-0_doping}(a).
For certain fillings in the small $t_{2}$ regime, the system prefers
to arrange the skyrmions in stripe patterns, as illustrated in Fig.$\,$\ref{fig:SUP_t2-0.025_doping}(b).
For $t_{2}=0$, there is a significant range of doping for which a
crystal of only $\chi=2$ skyrmions hosting two electrons each is formed,
as shown in Fig.$\,$\ref{fig:SUP_t2-0_doping}(b) (if the number
of electrons is odd, an additional skyrmion defect containing the
extra electron is formed). Depending on doping, these can also arrange
in stripe patterns, as exemplified in Fig.$\,$\ref{fig:SUP_t2-0_doping}(c). 

\paragraph*{}

\paragraph*{Electron density in DW phase with $V\protect\neq0$.--- }

\begin{figure}[H]
\begin{centering}
\includegraphics[width=0.55\columnwidth]{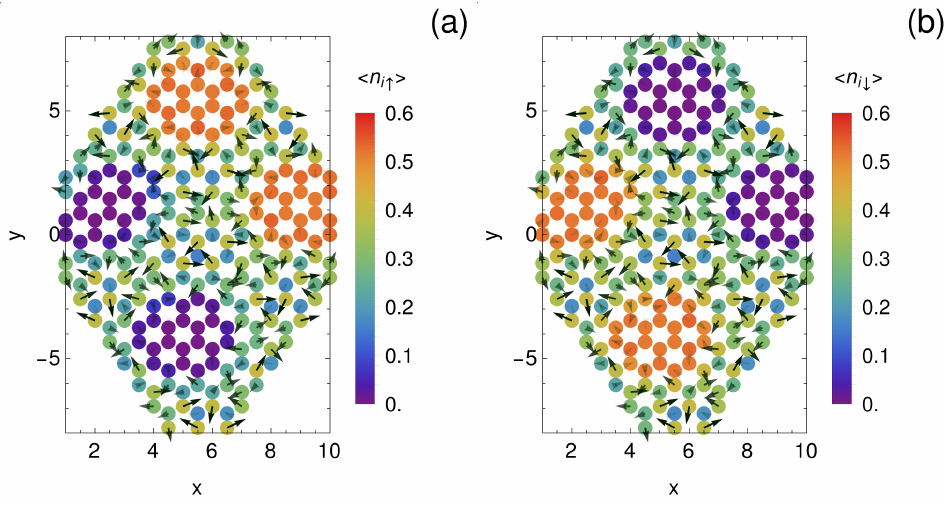}
\par\end{centering}
\caption{Electron density $\langle n_{i\uparrow}\rangle$ (a) and $\langle n_{i\downarrow}\rangle$
(b) within the DW phase at $V\protect\neq0$. Results for $V=1.05,U=8,t_{2}=0.2,L=12$
and $\delta_{e}=24$. \label{fig:SUP_DW_density}}

\end{figure}

We finish this section by providing results on the electron density
in the DW phase for $V\neq0$. As mentioned in the main text, a sublattice
density modulation is obtained in the coplanar domains, while the
ferromagnetic domains have uniform density. This is illustrated in
Fig.$\,$\ref{fig:SUP_DW_density}.

\section{Validity of skyrmion ansatz and skyrmion to polaron transition}

\label{sec:validity_ansatz}

\begin{figure}[H]
\begin{centering}
\includegraphics[width=1\columnwidth]{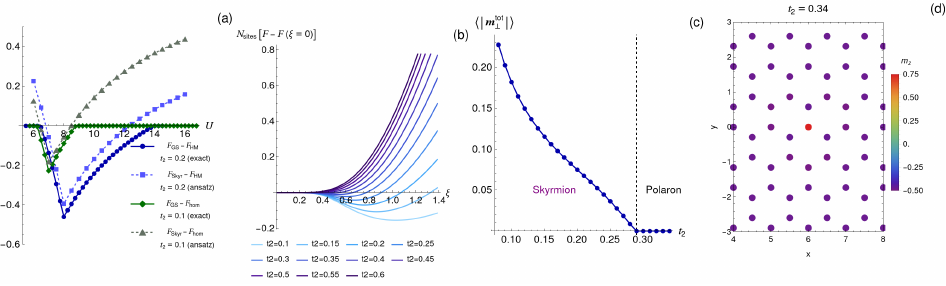}
\par\end{centering}
\caption{Validity of the skyrmion ansatz and skyrmion to polaron transition.
(a) Free energy results for $t_{2}=0.2,\delta_{e}=1,L=16$, for the
skyrmion ansatz (dashed) and the exact mean-field solution (full).
$F_{\textrm{GS}},F_{\textrm{Skyr}}$ and $F_{\textrm{HM}}$ stand
respectively for exact ground state free energy, ground state free
energy obtained through skyrmion ansatz and free energy of the HM
homogeneous solution. (b) Free energy difference between finite $\xi>0$
and $\xi=0$ using the skyrmion ansatz, for $U=8$ and $L=12$. For
$t_{2}\gtrsim0.35$, $\xi=0$ (polaron solution) minimizes the free
energy. (c) Results for the average coplanar magnetization
$\langle{\bf m}_{\perp}^{\textrm{tot}}\rangle=N_{\textrm{sites}}^{-1}\sum_{i}\sqrt{(m_{x}^{i})^{2}+(m_{y}^{i})^{2}}$,
for $U=8$, $L=12$ and $\delta_{e}=1$. For $t_{2}\protect\geq0.29$,
the coplanar magnetization vanishes, indicating that the ground state
becomes a polaron whose magnetization profile is exemplified in (d)
for $t_{2}=0.34$.\label{fig:SUP_validity_skyrmion_ansatz}}

\end{figure}

In this section we demonstrate the validity of the skyrmion ansatz
proposed in Eq.$\,$\ref{eq:ansatz} to describe the effective exchange
field spontaneously generated for $\delta_{e}=1$. In Fig.$\,$\ref{fig:SUP_validity_skyrmion_ansatz}(a)
we plot the difference in the free energy between the skyrmion ground state
and half-metal solutions to show that the regime of stability predicted
by the ansatz is qualitatively (and almost quantitatively) consistent
with the one predicted by solving exactly the mean-field equations. 

From the ansatz, it is also possible to predict that above a critical
$t_{2}$, the skyrmion ground state becomes unfavorable compared to
the polaron ground state (spin-flip magnetic impurity with null skyrmion
charge). This is explicitly shown in Fig.$\,$\ref{fig:SUP_validity_skyrmion_ansatz}(b).
The underlying reason is that both the magnitude of the topological
term and the skyrmion size $\xi$ decrease with $t_{2}$,
as shown in Fig.$\,$\ref{fig:2}(d) of the main text, which in turn
make the skyrmion texture become energetically unfavorable at large
enough $t_{2}$. These predictions are consistent with the exact result
shown in Fig.$\,$\ref{fig:SUP_validity_skyrmion_ansatz}(c). 

\section{Coplanar magnet at $\nu=4/3$}

\label{sec:SUP_coplanar_magnet}

\begin{figure}[H]

\begin{centering}
\includegraphics[width=0.3\columnwidth]{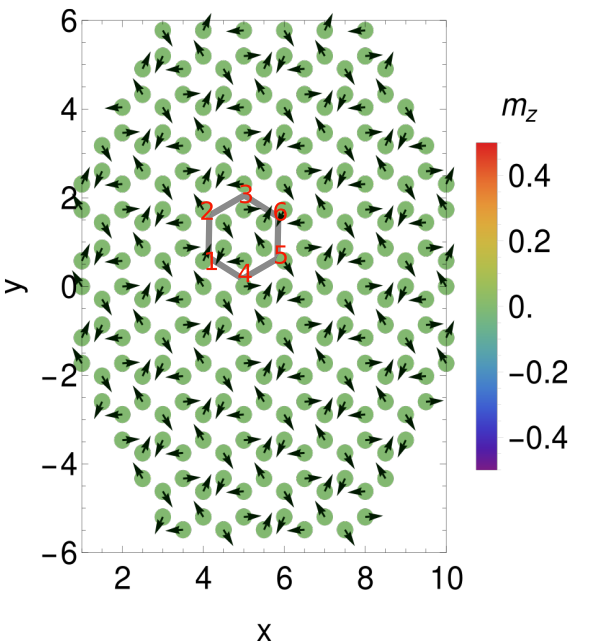}\caption{CoMI state at filling $\nu=4/3$. Unit cell and site ordering convention
used to derive the Hamiltonian in Eq.$\,$\ref{eq:MF_H_3UC} are sketched. \label{fig:SUP_site_numbering}}
\par\end{centering}
\end{figure}

Here we provide additional details on the calculations for the CoMI
phase at $\nu=4/3$. Using the unit cell definition and site numbering
in Fig.$\,$\ref{fig:SUP_site_numbering}, and defining 

\begin{equation}
f_{2}({\bf x},{\bf y})=it_{2}(1+e^{i{\bf x}\cdot{\bf k}}+e^{i{\bf y}\cdot{\bf k}}),
\end{equation}
the full Hamiltonian in ${\bf k}$-space is given by

\begin{equation}
H({\bf k})=\left(\begin{array}{cc}
H_{\up}({\bf k}) & H_{\up\down}({\bf k})\\
H_{\down\up}({\bf k}) & H_{\down}({\bf k})
\end{array}\right),\label{eq:MF_H_3UC}
\end{equation}
where

{\scriptsize{}
\begin{equation}
H_{\up}({\bf k})=\left(\begin{array}{cccccc}
0 & -t & f_{2}^{*}\Big({\bf a}'_{1},({\bf a}'_{1}-{\bf a}'_{2})\Big) & -t & -f_{2}^{*}\Big({\bf a}'_{1},{\bf a}'_{2}\Big) & -te^{-i{\bf a}'_{1}\cdot{\bf k}}\\
-t & 0 & -t & -f_{2}^{*}\Big({\bf a}'_{2},({\bf a}'_{2}-{\bf a}'_{1})\Big) & -te^{-i{\bf a}'_{2}\cdot{\bf k}} & f_{2}^{*}\Big({\bf a}'_{1},{\bf a}'_{2}\Big)\\
f_{2}\Big({\bf a}'_{1},({\bf a}'_{1}-{\bf a}'_{2})\Big) & -t & 0 & -te^{i({\bf a}'_{1}-{\bf a}'_{2})\cdot{\bf k}} & f_{2}^{*}\Big({\bf a}'_{2},({\bf a}'_{2}-{\bf a}'_{1})\Big) & -t\\
-t & -f_{2}\Big({\bf a}'_{2},({\bf a}'_{2}-{\bf a}'_{1})\Big) & -te^{i({\bf a}'_{2}-{\bf a}'_{1})\cdot{\bf k}} & 0 & -t & -f_{2}^{*}\Big({\bf a}'_{1},({\bf a}'_{1}-{\bf a}'_{2})\Big)\\
-f_{2}\Big({\bf a}'_{1},{\bf a}'_{2}\Big) & -te^{-i{\bf a}'_{2}\cdot{\bf k}} & f_{2}\Big({\bf a}'_{2},({\bf a}'_{2}-{\bf a}'_{1})\Big) & -t & 0 & -t\\
-te^{i{\bf a}'_{1}\cdot{\bf k}} & f_{2}\Big({\bf a}'_{1},{\bf a}'_{2}\Big) & -t & -f_{2}\Big({\bf a}'_{1},({\bf a}'_{1}-{\bf a}'_{2})\Big) & -t & 0
\end{array}\right)
\end{equation}
}{\scriptsize\par}

\begin{equation}
H_{\down}({\bf k})=H_{\up}({\bf k},t_{2}\rightarrow-t_{2})
\end{equation}

\begin{equation}
H_{\up\down}({\bf k})=MU\textrm{diag}(1,e^{i\theta},e^{2i\theta},e^{5i\theta},e^{4i\theta},e^{3i\theta}),\textrm{ \ensuremath{\theta=\pi/3}}
\end{equation}

\begin{equation}
H_{\down\up}({\bf k})=H_{\up\down}^{\dagger}({\bf k})
\end{equation}

\begin{equation}
\begin{array}{cc}
{\bf a}'_{1}= & (\frac{3}{2},\frac{\sqrt{3}}{2})\\
{\bf a}'_{2}= & (\frac{3}{2},-\frac{\sqrt{3}}{2})
\end{array}.
\end{equation}

We will now provide more details on the derivation of the continuum
model in the main text, Eq.$\,$\ref{eq:H_continuum}. As stated in
the main text, the Fourier transforms of $\langle c_{{\bf r}_{m},\up}^{\dagger}c_{{\bf r}_{m},\down}\rangle$
peak at ${\bf K}$ or ${\bf K}'$ depending on the sublattice and
on $w$. From this observation, we have $\langle c_{{\bf r}_{m},\sigma}^{\dagger}c_{{\bf r}_{m},-\sigma}\rangle=M\sum_{{\bf b}}e^{-iw\sigma m[\phi+({\bf K}+{\bf b})\cdot{\bf r}_{m}]}$,
where $M\propto U$, $m=\pm1$ respectively for sublattices $A$ and
$B$, ${\bf b}$ are the original reciprocal lattice vectors and
$\phi=\pi/6$ is the angle difference between sublattices. This yields
the following contribution to the mean-field Hamiltonian,

\begin{equation}
Me^{i\sigma wm\phi}\sum_{{\bf b},{\bf k}}c_{{\bf k},m,-\sigma}^{\dagger}c_{{\bf k}-\sigma wm({\bf K}+{\bf b}),m,\sigma}\,.
\end{equation}

Let us consider the first-order process in perturbation theory, taking
${\bf q}={\bf K},{\bf K}'$. For $w=-1$, we have

\begin{equation}
\begin{array}{cc}
Me^{-i\phi}\sum_{{\bf b}}c_{{\bf K}',-1,\down}^{\dagger}c_{{\bf K}'-({\bf K}+{\bf b}),-1,\up}+Me^{i\phi}\sum_{{\bf b}}c_{{\bf K},1,\down}^{\dagger}c_{{\bf K}+({\bf K}+{\bf b}),1,\up}\\
+Me^{-i\phi}\sum_{{\bf b}}c_{{\bf K}',1,\up}^{\dagger}c_{{\bf K}'-({\bf K}+{\bf b}),1,\down}+Me^{i\phi}\sum_{{\bf b}}c_{{\bf K},-1,\up}^{\dagger}c_{{\bf K}+({\bf K}+{\bf b}),1,\down}\\
=Me^{-i\phi}\sum_{{\bf b}}c_{{\bf K}',-1,\down}^{\dagger}c_{{\bf K}-{\bf b},-1,\up}+Me^{i\phi}\sum_{{\bf b}}c_{{\bf K},1,\down}^{\dagger}c_{{\bf K}'+{\bf b},1,\up}\\
+Me^{-i\phi}\sum_{{\bf b}}c_{{\bf K}',1,\up}^{\dagger}c_{{\bf K}-{\bf b},1,\down}+Me^{i\phi}\sum_{{\bf b}}c_{{\bf K},-1,\up}^{\dagger}c_{{\bf K}'+{\bf b},1,\down}
\end{array}
\end{equation}
while for $w=1$, we have

\begin{equation}
\begin{array}{cc}
Me^{i\phi}\sum_{{\bf b}}c_{{\bf K},-1,\down}^{\dagger}c_{{\bf K}+({\bf K}+{\bf b}),-1,\up}+Me^{-i\phi}\sum_{{\bf b}}c_{{\bf K}',1,\down}^{\dagger}c_{{\bf K}'-({\bf K}+{\bf b}),1,\up}\\
+Me^{i\phi}\sum_{{\bf b}}c_{{\bf K},1,\up}^{\dagger}c_{{\bf K}+({\bf K}+{\bf b}),1,\down}+Me^{-i\phi}\sum_{{\bf b}}c_{{\bf K}',-1,\up}^{\dagger}c_{{\bf K}'-({\bf K}+{\bf b}),1,\down}\\
=Me^{i\phi}\sum_{{\bf b}}c_{{\bf K},-1,\down}^{\dagger}c_{{\bf K}'+{\bf b},-1,\up}+Me^{-i\phi}\sum_{{\bf b}}c_{{\bf K}',1,\down}^{\dagger}c_{{\bf K}-{\bf b},1,\up}\\
Me^{i\phi}\sum_{{\bf b}}c_{{\bf K},1,\up}^{\dagger}c_{{\bf K}'+{\bf b},1,\down}+Me^{-i\phi}\sum_{{\bf b}}c_{{\bf K}',-1,\up}^{\dagger}c_{{\bf K}-{\bf b},1,\down}
\end{array}
\end{equation}

From these expressions, we can derive the low-energy continuum model
in Eq.$\,$\ref{eq:H_continuum}, neglecting the 2 degenerate lowest
and highest energy bands of the Hamiltonian in Eq.$\,$\ref{eq:MF_H_3UC}.
The energy bands for this model are given by 

\begin{equation}
\begin{array}{c}
E({\bf q})=\begin{cases}
-\frac{M}{2}\pm\sqrt{v_{f}^{2}{\bf q}^{2}+\Big(\frac{M}{2}-w\lambda_{\textrm{SO}}\Big)^{2}}\\
\frac{M}{2}\pm\sqrt{v_{f}^{2}{\bf q}^{2}+\Big(\frac{M}{2}+w\lambda_{\textrm{SO}}\Big)^{2}}
\end{cases}\end{array}
\end{equation}

Note that there are only four different dispersions even though there
are 2 flavours because each band is 2-fold degenerate, in agreement
with the exact model. The reason is that, as stated in the main text,
this Hamiltonian matrix can be written in terms of two identical $4\times4$
blocks if the basis elements are rearranged as 

\begin{equation}
\begin{array}{cc}
(\psi_{A,{\bf K},\up},\psi_{B,{\bf K},\up},\psi_{A,{\bf K}',\down},\psi_{B,{\bf K}',\down};\,\psi_{B,{\bf K},\down},\psi_{A,{\bf K},\down},\psi_{B,{\bf K}',\up},\psi_{A,{\bf K}',\up}).\end{array}
\end{equation}

\section{Observables}

\label{sec:SUP_Observables}

\subsection{Hall conductivity and Chern number}

\subsubsection{Kubo formula for tight-binding Hamiltonians}

We start with a general tight-binding Hamiltonian given by

\begin{equation}
H_{0}=\sum_{\bm{R}\bm{R}'\alpha\beta}t_{\bm{R}\bm{R}'}^{\alpha\beta}c_{\bm{R},\alpha}^{\dagger}c_{\bm{R}'\beta}
\end{equation}

We introduce the coupling to the vector potential to a Peierl's phase,
that is, $t_{\bm{R}\bm{R}'}^{\alpha\beta}\rightarrow t_{\bm{R}\bm{R}'}^{\alpha\beta}e^{-ie\int_{\bm{r}_{\alpha}}^{\bm{r}'_{\beta}}\bm{A}(\bm{r},t)\cdot d\bm{r}}$,
where $\bm{r}_{\alpha}$ is the position of site $\alpha$ belonging
to the unit cell $\bm{R}$. Within linear response, we expand to quadratic
terms in $\bm{A}$ and assume that it is constant over a lattice spacing
to get 

\begin{equation}
t_{\bm{R}\bm{R}'}^{\alpha\beta}\rightarrow t_{\bm{R}\bm{R}'}^{\alpha\beta}[1-ie\bm{A}\cdot\bm{\delta}+\frac{1}{2}e^{2}(\bm{A}\cdot\bm{\delta})^{2}]
\end{equation}

\noindent where $\bm{\delta}=\bm{r}'_{\beta}-\bm{r}_{\alpha}$. If
we now write $H=H_{0}+H'$, we have

\begin{equation}
H'(t)=-ie\sum_{\alpha\beta}\sum_{\bm{R}\bm{R}'}(\bm{A}(t)\cdot\bm{\delta})t_{\bm{R}\bm{R}'}^{\alpha\beta}c_{\bm{R},\alpha}^{\dagger}c_{\bm{R}'\beta}+\frac{1}{2}e^{2}\sum_{\alpha\beta}\sum_{\bm{R}\bm{R}'}(\bm{A}(t)\cdot\bm{\delta})^{2}t_{\bm{R}\bm{R}'}^{\alpha\beta}c_{\bm{R},\alpha}^{\dagger}c_{\bm{R}'\beta}
\end{equation}

or 

\begin{equation}
H'(t)=-j_{\mu}^{P}A_{\mu}(t)+\frac{1}{2}A_{\mu}(t)\Delta_{\mu\nu}A_{\nu}(t)\label{eq:H_prime_A_t}
\end{equation}

where the first and second terms are respectively the paramagnetic
and diamagnetic components of the current, with

\begin{equation}
j_{\mu}^{P}=ie\sum_{\alpha\beta}\sum_{\bm{R}\bm{R}'}t_{\bm{R}\bm{R}'}^{\alpha\beta}c_{\bm{R},\alpha}^{\dagger}c_{\bm{R}'\beta}\delta^{\mu}
\end{equation}

\begin{equation}
\Delta_{\mu\nu}=e^{2}\sum_{\alpha\beta}\sum_{\bm{R}\bm{R}'}t_{\bm{R}\bm{R}'}^{\alpha\beta}c_{\bm{R},\alpha}^{\dagger}c_{\bm{R}'\beta}\delta^{\mu}\delta^{\nu}
\end{equation}

Assuming the linear coupling $H'(t)=-\int d\bm{r}\bm{j}(\bm{r})\cdot\bm{A}(\bm{r},t)$,
the total current operator is then given by

\begin{equation}
j_{{\rm tot}}^{\mu}=-\frac{\delta H'}{\delta A_{\mu}}=j_{P}^{\mu}-\Delta_{\mu\nu}A_{\nu}+\mathcal{O}(A^{2})
\end{equation}

Using the Kubo formula, we have

\begin{equation}
\langle j_{{\rm tot}}^{\mu}\rangle(t)=-\langle\Delta^{\mu\nu}\rangle_{0}A^{\nu}(t)-\int dt'\Pi^{\mu\nu}(t,t')A^{\nu}(t')
\end{equation}
where $\langle\rangle_{0}$ is the average value taken for ${\bf A}=0$,
or in frequency space,

\begin{equation}
\langle j_{{\rm tot}}^{\mu}\rangle(\omega)=-[\langle\Delta^{\mu\nu}\rangle_{0}+\Pi^{\mu\nu}(\omega)]A^{\nu}(\omega)\,,\label{eq:total_current}
\end{equation}
where 

\begin{equation}
\Pi^{\mu\nu}(t,t')=-i\Theta(t-t')\langle[j_{P}^{\mu}(t),j_{P}^{\nu}(t')]\rangle_{0}\,,
\end{equation}
and $\Pi^{\mu\nu}(\omega)=\int_{0}^{+\infty}dte^{i\omega t}\Pi^{\mu\nu}(t,0)$.
Note that only ${\bf j}_{P}$ is considered in the commutator above
(and not ${\bf j}_{{\rm tot}}$) otherwise we would have a quadratic
contribution to $\langle j_{{\rm tot}}^{\mu}\rangle$. 

\subsubsection{Calculation of $\Pi^{\mu\nu}(\omega)$}

We will start by computing $\Pi^{\mu\nu}(t,t')$. First, we write
the Hamiltonian in the unperturbed eigenbasis basis, with $d_{n}^{\dagger}=\sum_{\bm{R}\alpha}a_{\bm{R}\alpha}^{n}c_{\bm{R}\alpha}^{\dagger}$
and $c_{\bm{R}\alpha}^{\dagger}=\sum_{n}(a_{\bm{R}\alpha}^{n})^{*}d_{n}^{\dagger}$,

\begin{equation}
H_{0}=\sum_{n}\epsilon_{n}d_{n}^{\dagger}d_{n}.
\end{equation}

We first compute $j_{\mu}^{P}(t)$ in this eigenbasis,

\begin{equation}
j_{\mu}^{P}(t)=\sum_{nm}j_{nm}^{\mu,P}d_{n}^{\dagger}(t)d_{m}(t)=\sum_{nm}j_{nm}^{\mu,P}e^{i(\epsilon_{n}-\epsilon_{m})t}d_{n}^{\dagger}d_{m}\,,
\end{equation}
where we used $d_{n}^{\dagger}(t)=e^{i\epsilon_{n}t}d_{n}^{\dagger}$
and

\begin{equation}
j_{nm}^{\mu,P}=ie\sum_{\alpha\beta}\sum_{\bm{R}\bm{R}'}t_{\bm{R}\bm{R}'}^{\alpha\beta}(a_{\bm{R}\alpha}^{n})^{*}a_{\bm{R}'\beta}^{m}\delta^{\mu}
\end{equation}

We now compute $\Pi^{\mu\nu}(t,t')$: 

\begin{equation}
\begin{aligned}\Pi^{\mu\nu}(t,t') & =-i\Theta(t-t')\sum_{nm}\sum_{ll'}j_{nm}^{\mu,P}e^{i(\epsilon_{n}-\epsilon_{m})t}j_{ll'}^{\nu,P}e^{i(\epsilon_{l}-\epsilon_{l'})t'}\langle[d_{n}^{\dagger}d_{m},d_{l}^{\dagger}d_{l'}]\rangle_{0}\\
 & =-i\Theta(t-t')\sum_{nm}j_{nm}^{\mu,P}j_{mn}^{\nu,P}e^{i(\epsilon_{n}-\epsilon_{m})(t-t')}[f(\epsilon_{n})-f(\epsilon_{m})]
\end{aligned}
\end{equation}

\noindent where we used $[d_{n}^{\dagger}d_{m},d_{l}^{\dagger}d_{l'}]=d_{n}^{\dagger}\{d_{m},d_{l}^{\dagger}\}d_{l'}-d_{l}^{\dagger}\{d_{n}^{\dagger},d_{l'}\}d_{m}=d_{n}^{\dagger}d_{l'}\delta_{ml}-d_{l}^{\dagger}d_{m}\delta_{nl'}$.
We will now work in frequency space. Using 

\begin{equation}
\Theta(t)=-\lim_{\eta\rightarrow0}\int\frac{d\omega}{2\pi i}\frac{e^{-i\omega t}}{\omega+i\eta}
\end{equation}
and 

\begin{equation}
\begin{aligned}\int\frac{d\omega'}{2\pi i}\frac{1}{\omega'+i\eta}\int_{0}^{+\infty}dte^{i(\omega-\omega'+\epsilon_{n}-\epsilon_{m})t}\\
=\frac{1}{i}\frac{1}{\omega+\epsilon_{n}-\epsilon_{m}+i\eta}\,,
\end{aligned}
\end{equation}
we get

\begin{equation}
\begin{aligned}\Pi^{\mu\nu}(\omega) & =\int_{0}^{+\infty}dte^{i\omega t}\Pi^{\mu\nu}(t,0)\\
 & =i\sum_{nm}j_{nm}^{\mu,P}j_{mn}^{\nu,P}[f(\epsilon_{n})-f(\epsilon_{m})]\int\frac{d\omega'}{2\pi i}\frac{1}{\omega'+i\eta}\int_{0}^{+\infty}dte^{i(\omega-\omega'+\epsilon_{n}-\epsilon_{m})t}\\
 & =\sum_{n\neq m}j_{nm}^{\mu,P}j_{mn}^{\nu,P}\frac{f(\epsilon_{n})-f(\epsilon_{m})}{\omega+\epsilon_{n}-\epsilon_{m}+i\eta}\,.
\end{aligned}
\end{equation}

\subsubsection{Calculation of $\langle\Delta^{\mu\nu}\rangle_{0}$}

We still need to evaluate $\langle\Delta^{\mu\nu}\rangle_{0}$. This
is the contribution from the diamagnetic part and therefore can be
computed using time-independent perturbation theory, by applying a
static vector potential $\bm{A}$. The total current $\langle j_{{\rm s,tot}}^{\mu}\rangle$
which is the average current after applying the static $\bm{A}$ is 

\begin{equation}
\langle j_{{\rm s,tot}}^{\mu}\rangle=\frac{1}{Z}\Tr[e^{-\beta H_{A}}j_{{\rm s,tot}}^{\mu}],
\end{equation}
where $j_{{\rm s,tot}}^{\mu}\equiv-\pd H_{A}/\pd A_{\mu}$. Similarly
to Eq.$\,$\ref{eq:H_prime_A_t}, we have 

\begin{equation}
H_{A}=-j_{s,P}^{\mu}A_{\mu}+\frac{1}{2}A_{\mu}\Delta_{\mu\nu}A_{\nu}
\end{equation}
where the superscript $s$ stands for static. By definition we have 

\begin{equation}
j_{s,P}^{\mu}=-\frac{\pd H_{A}}{\pd A_{\mu}}\Bigg|_{A=0}
\end{equation}

\begin{equation}
\Delta_{\mu\nu}^{s}=\frac{\pd^{2}H_{A}}{\pd A_{\mu}\pd A_{\nu}}\Bigg|_{A=0}
\end{equation}

To the lowest order, we have that $H_{A}=-\bm{j}_{s,P}\cdot\bm{A}+\mathcal{O}(A^{2})$
and $j_{{\rm s,tot}}^{\mu}=j_{s,P}^{\mu}-\Delta_{\mu\nu}A_{\nu}+\mathcal{O}(A^{2})$. 

Let us define $S(\tau)$ through

\begin{equation}
U(\tau)=e^{-\tau(H_{0}+V)}=e^{-\tau H_{0}}S(\tau)
\end{equation}
and from $\pd_{\tau}U(\tau)=-(H_{0}+V)U(\tau)$ we can derive

\begin{equation}
S(\tau)=\exp[-\int_{0}^{\tau}d\tau'V(\tau')]\approx1-\int_{0}^{\tau}d\tau'V(\tau')
\end{equation}

We now set $V(\tau)=-j_{P}^{\mu}(\tau)A_{\mu}$ and apply this result
to the thermal average $\langle j_{{\rm s,tot}}^{\mu}\rangle$ to
get, to order $\mathcal{O}(A^{2})$:

\begin{equation}
\begin{aligned}\langle j_{{\rm s,tot}}^{\mu}\rangle & =\frac{1}{Z}\Tr[e^{-\beta H_{A}}j_{{\rm s,tot}}^{\mu}]\\
 & \approx\frac{1}{Z}\Tr[e^{-\beta H_{0}}S(\beta)(j_{P}^{\mu}-\Delta_{\mu\nu}A_{\nu})]\\
 & =\langle j_{P}^{\mu}\rangle_{0}-\langle\Delta_{\mu\nu}\rangle_{0}A_{\nu}+\int_{0}^{\beta}d\tau'\langle T_{\tau}j_{P}^{\nu}(\tau')j_{P}^{\mu}\rangle_{0}A_{\nu}
\end{aligned}
\end{equation}

This implies that (noticing that $\langle j_{P}^{\mu}\rangle_{0}=0$):

\begin{equation}
\frac{\pd\langle j_{{\rm s,tot}}^{\mu}\rangle}{\pd A_{\nu}}\Bigg|_{A=0}=-\langle\Delta_{\mu\nu}\rangle_{0}+\int_{0}^{\beta}d\tau'\langle T_{\tau}j_{P}^{\nu}(\tau')j_{P}^{\mu}\rangle_{0}\label{eq:j_tot_and_diag}
\end{equation}

In the end we want to write $\langle\Delta_{\mu\nu}\rangle_{0}$ in
terms of the other two quantities. The second term is

\begin{equation}
\begin{aligned}\int_{0}^{\beta}d\tau'\langle j_{P}^{\mu}(\tau')j_{P}^{\nu}\rangle_{0} & =\int_{0}^{\beta}d\tau'\sum_{nm}\sum_{ll'}j_{nm}^{\mu,P}j_{ll'}^{\nu,P}\langle T_{\tau}d_{n}^{\dagger}(\tau')d_{m}(\tau')d_{l}^{\dagger}d_{l'}\rangle_{0}\\
 & =\int_{0}^{\beta}d\tau'\sum_{nm}\sum_{ll'}j_{nm}^{\mu,P}j_{ll'}^{\nu,P}[f(\epsilon_{n})f(\epsilon_{l})\delta_{mn}\delta_{ll'}-G_{n}(-\tau')G_{m}(\tau')\delta_{nl'}\delta_{ml}]
\end{aligned}
\label{eq:correlator_diamag}
\end{equation}
where $G_{\mu}(\tau)=-\langle T_{\tau}d_{\mu}(\tau)d_{\mu}^{\dagger}\rangle$.
The first term is proportional to

\begin{equation}
\sum_{n}j_{nn}^{\mu,P}f(\epsilon_{n})\sum_{l}j_{ll}^{\nu,P}f(\epsilon_{l})=\langle j_{P}^{\mu}\rangle_{0}\langle j_{P}^{\nu}\rangle_{0}=0
\end{equation}

For the second term in Eq.$\,$\ref{eq:correlator_diamag}, we work
in the Matsubara frequency space to get 

\begin{equation}
\begin{aligned}G_{\mu}(-\tau)G_{\nu}(\tau) & =\frac{1}{\beta^{2}}\sum_{\omega'_{n},\omega''_{n}}e^{i\tau(\omega'_{n}-\omega''_{n})}G_{\mu}(i\omega'_{n})G_{\nu}(i\omega''_{n})\end{aligned}
\end{equation}

\begin{equation}
\begin{aligned}\rightarrow\int_{0}^{\beta}d\tau'G_{\mu}(-\tau')G_{\nu}(\tau') & =\frac{1}{\beta}\sum_{\omega'_{n},\omega''_{n}}G_{\mu}(i\omega'_{n})G_{\nu}(i\omega''_{n})\frac{1}{\beta}\int_{0}^{\beta}d\tau'e^{i\tau(\omega'_{n}-\omega''_{n})}\\
 & =\frac{1}{\beta}\sum_{\omega'_{n}}G_{\mu}(i\omega'_{n})G_{\nu}(i\omega'_{n})
\end{aligned}
\end{equation}

The result can then be simplified. If $\mu\neq\nu$, we have

\begin{equation}
\begin{aligned}\frac{1}{\beta}\sum_{\omega'_{n}}G_{\mu}(i\omega'_{n})G_{\nu}(i\omega'_{n})=\frac{1}{\beta} & \sum_{\omega'_{n}}\frac{1}{i\omega'_{n}-\epsilon_{\mu}}\frac{1}{i\omega'_{n}-\epsilon_{\nu}}\\
=\frac{1}{2\pi i}\int dzf(z)\frac{1}{z-\epsilon_{\mu}}\frac{1}{z-\epsilon_{\nu}} & =\frac{f(\epsilon_{\mu})-f(\epsilon_{\nu})}{\epsilon_{\mu}-\epsilon_{\nu}}
\end{aligned}
\end{equation}

On the other hand, if $\epsilon_{\mu}=\epsilon_{\nu}$, we have

\begin{equation}
\begin{aligned}\frac{1}{\beta}\sum_{\omega'_{n}}G_{\mu}(i\omega'_{n})G_{\nu}(i\omega'_{n}) & =\frac{1}{2\pi i}\int dzf(z)\frac{1}{(z-\epsilon_{\mu})^{2}}=f'(\epsilon_{\mu})\end{aligned}
\end{equation}

\noindent where we used e ${\rm Res}(g,z_{0})=\frac{f^{(n-1)}(z_{0})}{(n-1)!}$
for $g(z)=\frac{f(z)}{(z-z_{0})^{n}}$. We therefore have

\begin{equation}
\int_{0}^{\beta}d\tau'\langle j_{s,P}^{\mu}(\tau')j_{s,P}^{\nu}\rangle_{0}=-\sum_{n\neq m}j_{nm}^{\mu,P}j_{mn}^{\nu,P}\frac{f(\epsilon_{n})-f(\epsilon_{m})}{\epsilon_{n}-\epsilon_{m}}-\sum_{\epsilon_{n}=\epsilon_{m}}j_{nm}^{\mu,P}j_{mn}^{\nu,P}f'(\epsilon_{n})
\end{equation}
and finally, from Eq.$\,$\ref{eq:j_tot_and_diag}

\begin{equation}
\langle\Delta_{\mu\nu}\rangle_{0}=-\frac{\pd\langle j_{{\rm s,tot}}^{\mu}\rangle}{\pd A_{\nu}}\Bigg|_{A=0}-\sum_{\epsilon_{n}=\epsilon_{m}}j_{nm}^{\mu,P}j_{mn}^{\nu,P}f'(\epsilon_{n})-\sum_{n\neq m}j_{nm}^{\mu,P}j_{mn}^{\nu,P}\frac{f(\epsilon_{n})-f(\epsilon_{m})}{\epsilon_{n}-\epsilon_{m}}
\end{equation}

In the absence of pairing terms, the first term should vanish. The
reason is that the total current due to an applied static potential,
$\langle j_{{\rm s,tot}}^{\mu}\rangle$, should vanish. Nonetheless,
we note that this term can be finite for finite systems. However,
it should vanish when the thermodynamic limit is taken.

\subsubsection{Combining $\Pi^{\mu\nu}(\omega)$ and $\langle\Delta_{\mu\nu}\rangle_{0}$
and computing the conductivity}

Recovering the expression

\begin{equation}
\langle j_{{\rm tot}}^{\mu}\rangle(\omega)=-[\langle\Delta^{\mu\nu}\rangle_{0}+\Pi^{\mu\nu}(\omega)]A^{\nu}(\omega)\,,
\end{equation}
we combine the last term of $\langle\Delta_{\mu\nu}\rangle_{0}$ with
$\Pi^{\mu\nu}(\omega)$ to get

\begin{equation}
\begin{aligned}\sum_{n\neq m}j_{nm}^{\mu,P}j_{mn}^{\nu,P}[f(\epsilon_{n})-f(\epsilon_{m})]\Big(\frac{1}{\omega+\epsilon_{n}-\epsilon_{m}+i\eta}-\frac{1}{\epsilon_{n}-\epsilon_{m}}\Big)\\
=-\sum_{n\neq m}j_{nm}^{\mu,P}j_{mn}^{\nu,P}\frac{f(\epsilon_{n})-f(\epsilon_{m})}{\epsilon_{n}-\epsilon_{m}}\frac{\omega+i\eta}{(\omega+\epsilon_{n}-\epsilon_{m}+i\eta)}
\end{aligned}
\end{equation}

The conductivity is given by 

\begin{equation}
\langle j_{{\rm tot}}^{\mu}\rangle(\omega)=\sigma_{\mu\nu}(\omega)E^{\nu}(\omega)
\end{equation}

Using that $E^{\nu}(\omega)=i(\omega+i\eta)A^{\nu}(\omega)$, we have
that 

\begin{equation}
\sigma_{\mu\nu}(\omega)=\frac{i}{\omega+i\eta}\Big[\langle\Delta^{\mu\nu}\rangle_{0}+\Pi^{\mu\nu}(\omega)\Big]
\end{equation}
and therefore

\begin{equation}
\sigma_{\mu\nu}(\omega)=-\frac{i}{\omega+i\eta}\Bigg[\frac{\pd\langle j_{{\rm s,tot}}^{\mu}\rangle}{\pd A_{\nu}}\Bigg|_{A=0}+\sum_{\epsilon_{n}=\epsilon_{m}}j_{nm}^{\mu,P}j_{mn}^{\nu,P}f'(\epsilon_{n})\Bigg]-i\sum_{n\neq m}\frac{f(\epsilon_{n})-f(\epsilon_{m})}{\epsilon_{n}-\epsilon_{m}}\frac{j_{nm}^{\mu,P}j_{mn}^{\nu,P}}{(\omega+\epsilon_{n}-\epsilon_{m}+i\eta)}\label{eq:conductivity_full}
\end{equation}

\subsubsection{Hall conductivity}

Taking the regular part of the Hall conductivity and setting $T=0$
in Eq.$\,$\ref{eq:conductivity_full}, we finally get

\begin{equation}
\sigma_{xy}=\sigma_{xy}(0)=-i\sum_{n\neq m}\frac{j_{nm}^{\mu,P}j_{mn}^{\nu,P}[f(\epsilon_{n})-f(\epsilon_{m})]}{(\epsilon_{n}-\epsilon_{m})^{2}}=\sum_{n\in\textrm{occ.},m\in\textrm{emp.}}\frac{2\Im[j_{nm}^{\mu,P}j_{mn}^{\nu,P}]}{(\epsilon_{n}-\epsilon_{m})^{2}}\label{eq:Hall_conductivity}
\end{equation}

\subsection{Skyrmion charge}

The skyrmion charge is given by 

\begin{equation}
\chi=\frac{1}{4\pi}\int d^{2}{\bf r}\,{\bf m}_{{\bf r}}\cdot(\pd_{x}{\bf m}_{{\bf r}}\times\pd_{y}{\bf m}_{{\bf r}}).
\end{equation}

This formula needs to be discretized in order to do the calculation
in the honeycomb lattice. A possible way introduced in Ref.$\,$\citep{Zhang2023}
is to map ${\bf m}$ to the spin-coherent state,

\begin{equation}
\ket{z_{j}}=\left(\begin{array}{c}
e^{-\frac{i\phi_{j}}{2}}\cos(\theta_{j}/2)\\
e^{\frac{i\phi_{j}}{2}}\sin(\theta_{j}/2)
\end{array}\right),
\end{equation}
where $j$ is the site index, such that 

\begin{equation}
\begin{array}{cc}
m_{x}^{j}=2\bra zS_{x}^{j}\ket z= & r_{j}\cos(\phi_{j})\sin(\theta_{j})\\
m_{y}^{j}=2\bra zS_{y}^{j}\ket z= & r_{j}\sin(\phi_{j})\sin(\theta_{j})\\
m_{z}^{j}=2\bra zS_{z}^{j}\ket z= & r_{j}\cos(\theta_{j})
\end{array}
\end{equation}

Then, we can compute $\chi$ by summing the berry phases over all
the hexagonal plaquettes in the system, as 

\begin{equation}
\chi=\sum_{\hexagon_{n}}\sum_{j\in\hexagon_{n}}\arg(\braket{\psi_{j}}{\psi_{j+\Delta}}).
\end{equation}

We refer to Ref.$\,$\citep{Zhang2023} for the complete derivation.

\section{Berry curvature distribution}

\label{sec:Berry-curvature-distribution}

In the main text, we established that even though a topological gap
is already present at filling $\nu=1$ for larger $t_{2}$, skyrmions
introduced upon doping play a crucial role for the quantization of
the Hall conductance, not simply acting as localized spectators in
the $\nu=1$ topological background. In fact, once skyrmions are
introduced as as in-gap states, the berry curvature re-distributes
and acquires a very significant weight in these states. Here we will
show additional data supporting these claims.

We examine how the berry curvature is distributed as a function
of energy. To do so, we write the Hall conductivity in Eq.$\,$\ref{eq:Hall_conductivity}
as

\begin{equation}
\sigma_{xy}=\sum_{n\in\textrm{occ.},m\in\textrm{emp.}}\frac{2\Im[j_{nm}^{\mu,P}j_{mn}^{\nu,P}]}{(\epsilon_{n}-\epsilon_{m})^{2}}=\sum_{n\in\textrm{occ.}}\Omega_{n}=\int dE\,\Omega(E)\,,
\end{equation}
where we defined the energy-dependent Berry curvature $\Omega(E)=\sum_{n\in\textrm{occ.}}\delta(E-\epsilon_{n})\,\tilde{\Omega}(E)$,
with $\tilde{\Omega}(E)=\sum_{m\in\textrm{emp.}}\frac{2\Im[j_{E,m}^{\mu,P}j_{m,E}^{\nu,P}]}{(E-\epsilon_{m})^{2}}$.
$\tilde{\Omega}(E)$ can be obtained by interpolation of the points
obtained at $E=\epsilon_{n}$. In practice, for the actual calculation,
we use a Lorentzian broadening for the Dirac delta function, approximating
$\delta(E-\epsilon_{n})\approx\eta^{2}/[(E-\epsilon_{n})^{2}+\eta^{2}]$,
and using $\eta\in[0.01-0.025]$. We also take $\tilde{\Omega}(E)\approx\tilde{\Omega}(E_{n})$
for the $n$-th term in the sum, given that the Lorentzian is very
peaked around $E=\epsilon_{n}$.

To complement these results, we will also compute the inverse participation
ratio for the mean-field single-particle eigenstates, defined as 

\begin{equation}
\textrm{IPR}_{n}=\frac{\sum_{j,\sigma}|\psi_{j,\sigma}^{n}|^{4}}{\big(\sum_{j,\sigma}|\psi_{j,\sigma}^{n}|^{2}\big)^{2}}
\end{equation}
for the $n$-th eigenstate $\ket{\psi_{n}}=\sum_{j,\sigma}\psi_{j,\sigma}^{n}\ket{j,\sigma}$,
with $j$ and $\sigma$ respectively the site and spin indices.

\begin{figure}[H]
\centering{}\includegraphics[width=1\columnwidth]{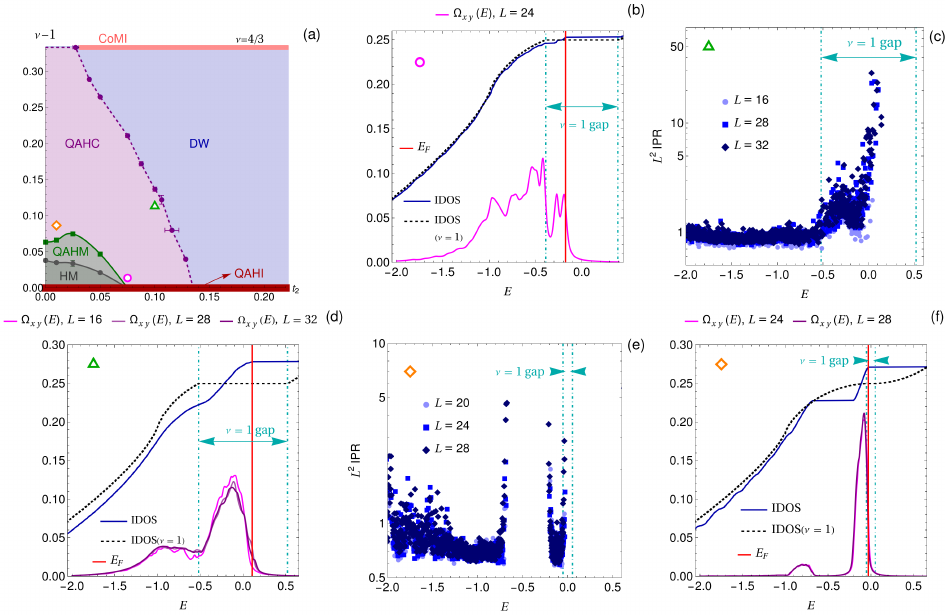}\caption{(a) Phase diagram for $U=8$, together with the points selected for
the plots in (b-f), shown in magenta, green and orange open markers.
(b) Integrated density of states (IDOS) and energy-resolved Berry
curvature $\Omega(E)$ for $t_{2}=0.075$ and $\nu\approx1.014$
{[}magenta point in panel (a){]}. In red we show the Fermi energy
and in cyan we delimit the topological gap that exists at $\nu=1$.
We also plot the IDOS for $\nu=1$ (dashed black line) for comparison.
(c) $L^{2}\textrm{IPR}$ for $t_{2}=0.1$ and $\nu\approx1.112$ {[}green
point in panel (a){]}. The data for different sizes has a good collapse
for all the shown eigenstates (up to the Fermi energy), indicating
that they are all extended. (d) IDOS and $\Omega(E)$ for the same
parameters as in panel (c). (e) $L^{2}\textrm{IPR}$ for $t_{2}=0.01$
and $\nu\approx1.088$ {[}orange point in panel (a){]}. (f) IDOS and
$\Omega(E)$ for the same parameters as in panel (e). \label{fig:OmegaE}}
\end{figure}

The results are shown in Fig.$\,$\ref{fig:OmegaE}. In short, they
indicate that skyrmions contribute to the Hall conductance even when
there is a sizable topological gap at $\nu=1$. More concretely,
when in-gap states are created upon doping, the Berry curvature is redistributed
and acquires a very significant contribution within these states.
In this way, skyrmions always contribute to the Hall conductance
at finite density above $\nu=1$ in the QAHC phase, with only the
fraction of the contribution changing for different parameters (larger
fraction for smaller $t_{2}$ and higher $\nu$).

The first example is shown in Fig.$\,$\ref{fig:OmegaE}(b), where
a very small doping is considered with respect to $\nu=1$. Because
the doping is small, only a tiny fraction of in-gap states is formed,
as indicated by the IDOS plot. Nonetheless, this small fraction of
states carries a significant contribution to $\sigma_{xy}$ as indicated
by the large values of $\Omega(E)$ inside the $\nu=1$ topological
gap, in Fig.$\,$\ref{fig:OmegaE}(b). 

For higher dopings, the fraction of states that occupies the $\nu=1$
gap gets larger and they start providing the most sizable contribution
to the Hall conductance, as shown in Fig.$\,$\ref{fig:OmegaE}(d), also shown in the main text.
This is consistent with the IPR results , where we observe $\textrm{IPR}\sim L^{-2}$
at finite density for all states, including those created inside the
$\nu=1$ gap, see Fig.$\,$\ref{fig:OmegaE}(c). This result implies
that the in-gap states are extended and can therefore contribute to
the Hall conductivity. 

For small $t_{2}$, the topological gap at $\nu=1$ is very small
and it no longer makes sense to interpret skyrmions as arising from
in-gap states upon doping. Instead, a sizable Chern gap is spontaneously
opened and by far the larger contributions for $\sigma_{xy}$ arise
for states close to the Fermi energy as shown in Fig.$\,$\ref{fig:OmegaE}(f).
This again indicates that skyrmions play a crucial role for the Hall
response, which is again consistent with the IPR results in Fig.$\,$\ref{fig:OmegaE}(e). 

\end{document}